\documentclass[compsoc,conference,a4paper,10pt,times]{IEEEtran}
\usepackage{newtxtext,newtxmath} 
\usepackage{url}
\usepackage{lipsum}
\usepackage{graphicx} 
\usepackage{subcaption}
\usepackage{listings}
\usepackage{textcomp}
\usepackage{xcolor}
\usepackage{acronym}
\usepackage{comment}
\usepackage{enumitem} 
\usepackage{hyperref}
\hypersetup{
    colorlinks=true, 
    linkcolor=black,  
    citecolor=black, 
    urlcolor=black,   
    filecolor=black, 
    bookmarks=true, 
    bookmarksnumbered=true, 
    linktocpage=true 
}

\usepackage{booktabs}
\usepackage{tabularx}
\usepackage{multirow}
\usepackage{siunitx}
\usepackage{float}
\usepackage{adjustbox}
\usepackage{minted}
\usepackage{cleveref}
\usepackage{caption} 
\usepackage{subcaption} 
\usepackage{booktabs}

\crefname{section}{§}{§§}
\crefname{subsection}{§}{§§}
\crefname{figure}{Fig.}{Figs.}
\crefname{table}{Tab.}{Tabs.}
\crefname{listing}{Listing}{Listings}

\begin{document}
\title{
SCyTAG: Scalable Cyber-Twin for Threat-Assessment Based on Attack Graphs
}

\author{
    David~Tayouri$^1$,
    {Elad~Duani$^1$},
    {Abed~Showgan$^1$},
    {Ofir~Manor$^2$},
    {Ortal~Lavi$^2$},
    {Igor~Podoski$^3$},
    {Miro~Ohana$^1$},\\
    {Yuval~Elovici$^1$},
    {Andres~Murillo$^2$},
    {Asaf~Shabtai$^1$},
    {Rami~Puzis$^1$}\\

    \small
    $^1$ Ben-Gurion University of the Negev,
    $^2$ Fujitsu Research Europe,
    $^3$ Fujitsu Technology Solutions\\

    \scriptsize
    \{davidtay,duani,miromeir\}@post.bgu.ac.il, showganabed@gmail.com, \{ofir.manor,ortal.lavi,igor.podoski,andres.murillo\}@fujitsu.com, \{elovici,shabtaia,puzis\}@bgu.ac.il
    
}

\maketitle

\begin{abstract}

Understanding the risks associated with an enterprise environment is the first step toward improving its security.
Organizations employ various methods to assess and prioritize the risks identified in \ac{CTI} reports that may be relevant to their operations. 
Some methodologies rely heavily on manual analysis (which requires expertise and cannot be applied frequently), while others automate the assessment, using \acp{AG} or threat emulators.
Such emulators can be employed in conjunction with cyber twins to avoid disruptions in live production environments when evaluating the highlighted threats. 
Unfortunately, the use of cyber twins in organizational networks is limited due to their inability to scale.  
In this paper, we propose \textit{SCyTAG}, a multi-step framework that generates the minimal viable cyber twin required to assess the impact of a given attack scenario. 
Given the organizational computer network specifications and an attack scenario extracted from a \ac{CTI} report, \textit{SCyTAG} generates an \ac{AG}. 
Then, based on the \ac{AG}, it automatically constructs a cyber twin comprising the network components necessary to emulate the attack scenario and assess the relevance and risks of the attack to the organization.
We evaluate \textit{SCyTAG} on both a real and fictitious organizational network. 
The results show that compared to the full topology, \textit{SCyTAG} reduces the number of network components needed for emulation by up to 85\% and halves the amount of required resources while preserving the fidelity of the emulated attack.
\textit{SCyTAG} serves as a cost-effective, scalable, and highly adaptable threat assessment solution, improving organizational cyber defense by bridging the gap between abstract \ac{CTI} and practical scenario-driven testing.
\end{abstract}


\section{Introduction}
\label{sec:introduction}
The increased sharing of information about cyber threats seen in recent years has improved the preparedness of organizations for the onset of attack campaigns.
\Acf{CTI} reports shared by threat advisories and security companies contain information about threat actors' \acp{TTP} and other information that facilitates attack detection. 
\Ac{CTI} reports enable organizations to conduct a timely threat assessment, consider the risks, and plan defenses.  
However, determining how an attack described in a \ac{CTI} report would manifest itself in a given enterprise environment is not a trivial task. 

Typically, security specialists identify, quantify, and prioritize risks based on the latest \ac{CTI}, known vulnerabilities, organizational security posture, and more~\cite{CybersecurityRiskAssessment,ThreatAssessment}.  
Various approaches have been proposed to assess the potential impact of an anticipated cyberattack and explore mitigations, including the use of \acp{DT}~\cite{oyama2023digital} (see ~\cref{subsec:DigitalTwins}).
In this paper, we use the term cyber twin to refer to a \ac{DT} that is used to emulate cyberattacks.
Cyber twins, which provide a means of safely examining the impact of attacks, launching exploits, and observing consequences without risking production services, enable fine-grained threat assessment.
However, scalability is a major challenge when employing cyber twins in enterprise networks, which are large and heterogeneous in terms of hosts, devices, and configurations.
Generating a cyber twin for such a network is either not feasible or very expensive.
\Acfp{AG}~\cite{sembiring2015network,stergiopoulos2022automatic} (see~\cref{subsec:AgAndMulval}) are also used in threat assessment, by enabling analysts to enumerate the possible ways for an attacker to infiltrate a network, obtain assets, and achieve goals. 
\Acp{AG} are more scalable than \acp{DT}, but they lack the ability to emulate attacks and test defenses.
The research questions we want to answer are:
\begin{itemize}
    \item How to evaluate the impact of a given attack scenario on a large organization's network?
    \item How to find the minimal set of network entities that are sufficient for building a \ac{DT} of the network and emulating the attack?
    \item How can we build a scalable and efficient process that can be run periodically and automatically for different attack scenarios and in different network environments?
\end{itemize}

In this paper, we introduce SCyTAG - Scalable Cyber-Twin for Threat-Assessment based on Attack Graphs.
Our proposed framework combines the \ac{AG} and cyber twin approaches to perform scalable fine-grained threat assessment (see \cref{sec:Methodology}).
First, the network entities relevant to an attack scenario are identified using an \ac{AG}, and then the attack is emulated in a cyber twin.
SCyTAG enables security teams to evaluate the resilience of their enterprise network to a concrete attack scenario.

The process starts by collecting the network topology, software configurations, and known vulnerabilities of the target network.
Then, for a given \ac{CTI}, SCyTAG generates an \ac{AG} with potential attack paths.
Based on the generated \ac{AG}, the framework constructs a nominal cyber twin: a minimal set of machines, services, and firewall rules required to emulate the entire attack scenario. 
By focusing only on essential components, this approach significantly reduces hardware needs and complexity while retaining enough details for a credible, network-specific test.
The cyber twin is then used to emulate the actual attack steps, executing each malicious action in the threat scenario. 
Security teams can also use SCyTAG for what-if analysis, and after employing defense mechanisms, the team can rerun the emulation to validate that all the vulnerabilities enabling the attack have been addressed.
The cyber twins generated by SCyTAG are also helpful for security training, playbook validation, and supply-chain vendor audits (see~\cref{subsec:UseCases}).

We demonstrate the SCyTAG framework on two networks: one closely follows the structure and configuration of a real enterprise, and the other is a complex fictitious network (see \cref{sec:Evaluation}).
Our evaluation shows that in comparison to the full topology, SCyTAG reduces the number of entities for the cyber twin generation by up to 85\%, 
without compromising the comprehensiveness and accuracy of the emulation.

\noindent The main contributions of this paper are:
\begin{itemize}[leftmargin=*]
    \item A framework that uses \acp{AG} and cyber twins to facilitate accurate and scalable risk assessment and validate network resilience to a variety of attack scenarios.
    \item Automatic generation of \ac{AG} facts from network topology and configuration files.
    \item Automatic generation of a minimal \ac{DT} that possesses all of the entities required for attack emulation.
    \item Automatic filling of Caldera Ability variables from \ac{AG} entities, for accurate attack emulation.
    \item Automatic execution of \ac{CTI}-derived attack on the generated cyber twin, enabling end-to-end impact assessment.
    \item Comprehensive evaluation of the proposed framework with one real (medium-size) and one fictitious (large-size) organizational network topologies.
\end{itemize}

\section{Background} \label{sec:Background}

\subsection{MITRE ATT\&CK, CTI, Attack Emulation} \label{subsec:AttackAndCti}
\subsubsection{MITRE ATT\&CK}
\Acfp{TTP} denote the concrete goal (tactic), way of action (technique), and specific behaviors (procedure) that adversaries employ for a cyberattack.
MITRE ATT\&CK (Adversarial Tactics, Techniques, and Common Knowledge) is a knowledge base of adversarial tactics and techniques based on real-world observations~\cite{strom2018mitre}.
The ATT\&CK knowledge base serves as a foundation for the development of threat models and methodologies by the cybersecurity community.
ATT\&CK provides a common taxonomy for both offense and defense and has become a useful conceptual tool in many cybersecurity disciplines, helping improve network and system defenses against intrusions.

Representing an attack in terms of Tactics, Techniques, and Sub-techniques provides a means of balancing the technical details in the Technique description and the attack goals represented by the Tactics.
Tactics represent the adversarial tactical objective, the "why" of an ATT\&CK Technique or Sub-technique. Meanwhile, Techniques represent the “how” - the actions an adversary performs to achieve that tactical objective.
Sub-techniques further break down the behaviors described by Techniques.
Procedures are the specific implementations that adversaries use to apply Techniques or Sub-techniques.
In addition to textual descriptions, metadata, Sub-techniques, and Procedures, a Technique may also include Group, Software, Mitigation, and Detection.

\subsubsection{Cyber Threat Intelligence}
\ac{CTI} reports contain knowledge about threat actors, their motives, targets, and ability to perform attacks.
\ac{CTI} about emerging threats and adversaries contributes to preparedness and the effective development of cybersecurity solutions~\cite{tounsi2018survey}.
\ac{CTI} can be divided into two primary categories: structured (such as STIX~\cite{barnum2012standardizing}, TAXII~\cite{connolly2014trusted}, and YARA~\cite{yara}) and unstructured.
Threat actors' \ac{TTP} are essential components of \ac{CTI}.
Some cybersecurity firms manually map their \ac{CTI} reports to MITRE ATT\&CK Techniques.
Several studies have proposed methods for the automatic extraction of MITRE ATT\&CK Techniques from \ac{CTI} reports, e.g.,~\cite{huang2024mitretrieval,legoy2020automated,liu2022threat}.

\subsubsection{Attack Emulation}
Attack (or threat) emulation is a cybersecurity assessment method used to test an organization's security controls against the \acp{TTP} used by threat actors posing risks to the organization.
This method enables organizations to emulate attack scenarios in a controlled environment, allowing them to assess their security posture and identify weaknesses.
The data obtained through attack emulation can also be used to track progress and shape future cybersecurity strategies, thereby improving an organization’s security posture.

MITRE Caldera~\cite{caldera} is an open-source, automated attack emulation platform that uses the MITRE ATT\&CK framework to model threats and replicate their behaviors.
With Caldera, cyber teams can create a customized cyberattack profile and deploy it within a network to identify vulnerabilities.
Caldera enables automated testing of cyber defenses.
It also supports the integration of network and host defenses, logging, sensors, analytics, alerts, and automated response.
Caldera enables automated offensive and defensive cyber operations, as well as cyber defense analytics.

\subsection{Digital Twins} \label{subsec:DigitalTwins}
\Acf{DT} is a set of virtual information constructs that fully describe a potential or actual physical system~\cite{grieves2017digital}.
Operationally, a \ac{DT} functions as a synchronized replica of its source environment, physical or purely virtual, continually ingesting telemetry and mirroring state changes. 
Conceived during the early 2000s, and in NASA product life-cycle studies \cite{unknown}, the \ac{DT} concept has matured rapidly in civil infrastructure and manufacturing applications over the past decade~\cite{liu2023literature}. 


A \ac{DT} must satisfy five interdependent requirements~\cite{feng2024game}:
\begin{enumerate}[label=(\roman*)]
\item Real‑time data synchronization - must ingest low-latency, distributed telemetry so that the replica continuously reflects the current network state.
\item Full lifecycle modeling - the model must span deployment, monitoring, diagnosis, and restoration phases so that failures, patches, and blue‑team responses can be rehearsed end‑to‑end.
\item High‑fidelity visualization - must provide interactive three‑dimensional or layered views to help analysts trace attack flows and verify mitigation steps.
\item Robust data services - must provide a hybrid store combining streaming access with historical archives, which supports both immediate response and forensic rollback, aligns with data‑infrastructure recommendations in \cite{liu2023literature}.
\item Standardized interfaces - must expose open \acp{API} so that simulation orchestrators, attack-graph engines, and other security tools can integrate directly into the twin runtime and drive complex scenario orchestration.
\end{enumerate}

\textbf{From Digital to Cyber Twins:} In cybersecurity, the emphasis shifts to network topologies, configurations, and adversarial behavior, yielding a safe environment where exploits can be launched, observed, and rolled back without jeopardizing production services.
Unlike general \acp{DT}, whose fidelity centers on mechanical geometry and sensor read‑outs, a \emph{cyber twin} reproduces the logical structure of an information system, including routing tables, firewall policies, host software stacks, and live traffic patterns.
This shift in focus is critical for security analysis because threats emerge from control‑flow dependencies rather than physical wear.
This logical-layer abstraction enables near real‑time replay of attack scenarios and validation of counter‑measures.

While usage is still emerging, following \cite{yu2019cybertwin} and subsequent studies like \cite{yigit2024cyber} and \cite{voas2025security}, we adopt the cyber twin term to refer to a \ac{DT} aimed at emulating cyberattacks.
By fulfilling the five requirements mentioned above, a static network model can be transformed into a living cyber twin that supports safe, repeatable emulation of cyber threats, rapid \ac{AG} regeneration, and automated response evaluation. 

\subsection{Attack Graphs and MulVAL} \label{subsec:AgAndMulval}
An \ac{AG} is a model that enables researchers and security professionals to enumerate all the attack paths and visually represent events that can result in a successful attack.
\Acp{AG} provide a formal representation of all of the exploit sequences an adversary can follow to reach a goal, such as a protected asset.


In this research, we utilize MulVAL, an open-source logic-based \ac{AG} generation tool~\cite{ou2005logic}. 
MulVAL is based on the Datalog modeling language, a subset of the Prolog logic programming language. 
In MulVAL, Datalog is used to represent two types of entities:
\begin{itemize}
\item \textit{Facts}: network topologies and configurations, security policies, and known vulnerabilities.
\item \textit{Rules} (also known as interaction rules): interactions between components in the network.
\end{itemize}

A sentence in MulVAL is defined as a clause of literals:
$$L_0 :- L_1,...,L_n$$ \label{def:IR}
\noindent
where $L_0$ is defined as the head, and $L_1,...,L_n$ comprise the body of the sentence.
Each $L_i$ in the body can be either a fact or \ac{IR}.
Body literals $(L_1,...,L_n)$ are preconditions for the head ($L_0$): if the body literals are true, then the head literal is also true.
A sentence with an empty body is called a fact.
A sentence with a nonempty body is called a rule.

As a \ac{LAG} generator, MulVAL's rules can be extended to represent known \acp{TTP}, making it suitable for modeling a wide range of threat scenarios characterized by attackers' goals, capabilities, and resources.
Our selection of MulVAL for this research is based on its documented advantages over other \ac{AG} generation tools. 

MulVAL has the advantage of extensibility: its underlying reasoning engine is written in a logical programming language, which enables extending functionality by writing custom rules.
We leverage this capability and define new \acp{IR} in order to generate \acp{AG} for attack scenarios that haven't yet been modeled.

\section{Methodology} \label{sec:Methodology}
\begin{figure*}[htbp]
    \centering
    \includegraphics[width=0.93\textwidth]{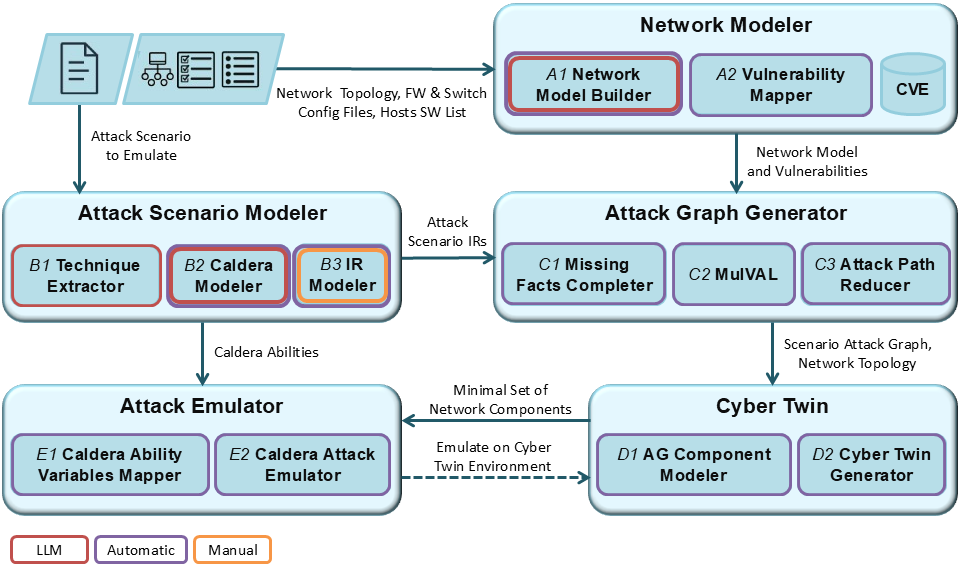}
    \caption{SCyTAG architecture.}
    \label{fig:SCyTAGArchitecture}
\end{figure*}

To assess the threat described in a \ac{CTI} report to a given organizational network, SCyTAG generates a small cyber twin of the organization that can emulate the described attack. 
The process begins with the analysis of the organization’s \ac{IT} specifications and the \ac{CTI} report.  
Then SCyTAG generates an \ac{AG} well aligned with techniques supported by the threat emulator deployed in the organization. 
In this paper, we consider the abilities supported by Caldera.  
Relying on the \ac{AG}, SCyTAG automatically generates a minimal viable cyber twin containing: \textbf{only} the \ac{IT} components (minimal) and \textbf{all} the \ac{IT} components (viable) required to estimate the attack impact on the organization. 
Finally, a threat emulator is used to assess the attack impact on the cyber twin.
SCyTAG's architecture, as presented in \cref{fig:SCyTAGArchitecture}, consists of five modules, which are elaborated on in the following subsections. 

\subsection{Network Modeler} \label{subsec:NetworkModeler}
The Network Modeler converts raw \ac{IT} specifications and vulnerabilities into logical facts.

\subsubsection{Network Model Builder} \label{subsec:NetworkModeleBuilder}
The inputs to this component are:
\begin{itemize}
\item \textbf{Network topology} - either text or diagram (in JPEG, PNG, SVG, or Visio formats) depicting elements and links.
\item \textbf{Software inventories} - the list of software installed on the network elements (received as JSON or CSV).
\item \textbf{Network configurations} - the configurations of communication elements (e.g., switches, routers, and firewalls).
\end{itemize}

If the topology is received in the form of diagrams, they are transformed to a textual format with the GeNet framework~\cite{ifland2024genet}.  
GeNet employs an \ac{LLM} that can perform visual analysis (GPT-4 Vision), with a purpose‑built prompt to: (i) enumerate diagram objects, (ii) classify device roles, (iii) list interface labels, and (iv) infer pairwise links.
Postprocessing scripts transform the LLM output into JSON nodes and edges.

For unstructured textual input, SCyTAG applies lightweight \ac{NLP}, whereas structured files are validated against JSON schema.
All artifacts are merged into a unified JSON schema that expresses nodes, interfaces, addressing, routing, software, and \acp{ACL}.

From the schema, SCyTAG generates MulVAL facts.
\cref{lst:BasicFacts} presents the facts used to represent the network model.
\texttt{\small{dataBind}} maps a data symbol to its physical location in the network.
\texttt{\small{dataFlow}} defines a connection between two network elements.
\texttt{\small{isInSubnet}} identifies a node's subnet.
\texttt{\small{belongTo}} and \texttt{\small{hasIP}} define subnet and network relationships.
\texttt{\small{networkService}} defines a service running on a host.
\texttt{\small{hacl}} defines a host access control policy.
\texttt{\small{residesOn}} defines a software running on a host.

\begin{lstlisting}[basicstyle=\ttfamily\scriptsize,language=Python,label=lst:BasicFacts, frame=single,caption=Topology facts]
dataBind(Flow, SrcHost, Path).
dataFlow(Host1, Host2, FlowName, Direction).
isInSubnet(Subnet, Host).
belongsTo(Subnet, VirtualPort).
hasIP(IP, Host).
networkService(Host, Software, Protocol, Port, Account).
hacl(SrcHost, DstHost, Protocol, DestPort).
residesOn(Host, Software, Version).
\end{lstlisting}

The output of this component is the list of facts representing the network.

\subsubsection{Vulnerability Mapper}
The input to this component is either the list of vulnerabilities found in the network elements or the list of software installed on these elements.
In the second case, the list of software is used to search in \acp{CVE}~\cite{cve} and extract the relevant known vulnerabilities of the software.
In both cases, \texttt{\small{vulExists}} (see \cref{lst:VulnerabilityFact}) is used to represent the vulnerabilities.

\begin{lstlisting}[basicstyle=\ttfamily\scriptsize, language=Python, label=lst:VulnerabilityFact, frame=single, caption=Vulnerability fact]
vulExists(CveId,SW,Version,AccessVec,LoseTypes,Severity).
\end{lstlisting}

The output of this component is the list of facts representing known vulnerabilities in the network's software.

The output of the Network Modeler module is the network model and its associated vulnerabilities, represented as MulVAL facts, which are then passed to the Attack Graph Generator module.

\subsection{Attack Scenario Modeler} \label{subsec:AttackScenarioModeler}
The Attack Scenario Modeler parses a given attack scenario, extracts ATT\&CK Techniques, and maps them to Caldera Abilities and MulVAL \acp{IR}.
The reason for choosing ATT\&CK for \ac{TTP} taxonomy is that it helps map the attack scenario to (Caldera) attack emulation and to modeled MulVAL \acp{IR}.

\subsubsection{Technique Extractor}
The input to the Technique Extractor component is an attack scenario, e.g., a \ac{CTI} report.
This report contains all of the attack information, such as the reporting agency, possible targets, threat group, and conclusion.
The report is input to the component's \ac{LLM}, along with the instructions to extract an ordered list of ATT\&CK Technique (and Sub-technique) IDs, which represent the attack steps.
With the \ac{LLM}, we employ a simple \ac{RAG} mechanism that uses the MITRE ATT\&CK knowledge database to match expressions in the \ac{CTI} report to descriptions of Tactics and Techniques.
The next step in the process is to verify that the \ac{LLM}'s output is an ordered list of Techniques; if the verification fails, the process is performed again.
The next step is to validate the output, ensuring that the selected IDs and their placement in the list are correct. If the validation fails, the process is repeated. 
The output is a verified and validated ordered list of Techniques.

\subsubsection{Caldera Modeler}
In Caldera, each attack step is called an Ability, and a set of Abilities representing an attack is called an Adversary.
The Caldera Modeler component's input is an ordered list of Techniques received from the Technique Extractor component.
This component uses the IDs of Techniques and Sub-techniques to map them to the available Caldera Abilities (YAML-defined agent operations for emulated execution).
When multiple abilities are found, we use the \ac{LLM} to perform similarity analysis on the attack pattern description to obtain the closest result and map each Technique to an Ability.
Since the list of Abilities may not cover a full attack scenario, the component tries to match the Ability list to an Adversary.
Missing Abilities can be completed manually.
The list of Abilities from the found Adversary is passed to the Attack Emulator module, and is also mapped back to Techniques.
This list of Techniques is passed to the \ac{IR} Modeler component.

\subsubsection{IR Modeler} \label{subsec:IRModeler}
The input to this component is the list of ATT\&CK Techniques received from the Caldera Modeler component.
Tayouri et al.~\cite{tayouri2023survey} mapped all of the \acp{IR} defined by researchers as of December 2022 to ATT\&CK Techniques.
Based on this mapping~\cite{mulval2mitre}, the \ac{IR} Modeler finds the relevant \acp{IR} for the given Techniques.
Techniques that have not yet been modeled to \acp{IR} should be modeled manually.
The output of the \ac{IR} Modeler component is a list of \acp{IR}, which are passed to the Attack Graph Generator module.

\subsection{Attack Graph Generator} \label{subsec:AttackGraphGenerator}
The Attack Graph Generator module runs MulVAL with the given facts and \acp{IR} to generate an \ac{AG} that represents the attack scenario to emulate.

\subsubsection{Missing Facts Completer} \label{subsec:MissingFactCompletor}
There may be facts that are conditions to the \acp{IR} representing the attack scenario and cannot be part of the organization’s \ac{IT} specifications, e.g., \texttt{\small{malicious(attacker)}} or \texttt{\small{incompetent(user)}}.
Such facts will cause immediate failure in the AG generation process.
This component processes the \acfp{IR} received from the \ac{IR} Modeler and the facts received from the Network Modeler, and completes the facts that are not part of \cref{lst:BasicFacts}, as they cannot be included in the network model.
This enables the network model facts to affect the AG generation for the given scenario.

\subsubsection{MulVAL}
In this component, MulVAL is run, based on the facts received from the Network Modeler, the additional facts from the Missing Facts Completer, and the set of \acfp{IR} received from the Attack Scenario Modeler.
It generates an \ac{AG} containing all of the possible exploitation paths for the given attack scenario, represented by the \acp{IR}.
This \ac{AG} serves as input to the Attack Path Reducer component.

\subsubsection{Attack Path Reducer}
The \ac{AG} produced by the MulVAL component may contain parallel or overlapping sequences of steps.
\ac{LAG} attack paths may be redundant, for example, attack paths that have the same final goal but differ in terms of vulnerabilities or intermediate hosts.
This component applies a reduction step, suggested by Gonda et al.~\cite{gonda2017scalable}, which removes unproductive branches while retaining the scenario’s genuinely feasible chain(s).

The suggested approach consists of two steps, which are performed to remove possible redundancy.
Merging Equivalent Outcomes: In situations where multiple exploit sequences yield the same result, e.g., “attacker privilege on Host X,” only one node is retained, ensuring that there is no duplication in the final graph.
Trimming Redundant Steps: Paths that replicate the same essential pivot or rely on the same exploit are unified.
If the scenario explicitly prescribes a single route for the pivot, alternative discovered paths can be omitted from the final output.

For example, all of the attack paths that achieve the same final goal but with different CVEs or via different hosts are reduced to a single path by (randomly) choosing a single host and a single CVE.
This reduction minimizes complexity, resource demands, and operational overhead, while preserving the attack paths necessary for emulating the given attack scenario.

The \ac{AG}, after duplicate paths are discarded, is the input for the Cyber Twin module.

\subsection{Cyber Twin}\label{subsec:CyberTwin}
The Cyber Twin module uses the \ac{AG} received from the Attack Graph Generator to create a cyber twin - a runnable, scenario‑focused virtual network in GNS3. 
Its guiding principle is to preserve \emph{every} host, service, and data‑flow edge that appears on at least one path of the \ac{AG}, maintaining full scenario fidelity.
Network entities that do not play a role in the attack are not part of the \ac{AG} and therefore are not part of the cyber twin.
By constructing a semantically faithful yet resource-efficient replica, the Cyber Twin module ensures that the cyber twin contains all of the network entities and relationships required for emulating the attack scenario.
This module ensures that every emulated step corresponds to a reachable path in the enterprise network.

\subsubsection{AG Component Modeler} \label{subsec:AGComponents}
This component receives the \ac{AG} and the whole network topology from the Attack Graph Generator.
The component extracts from the \ac{AG} the hosts, services, and other network assets, and verifies they are sufficient for generating a valid cyber twin.  
\acp{LAG} can drop intermediate infrastructure devices (e.g., core switches or distribution routers) if they are not part of any attack path.
To guarantee a runnable twin, we compare the \ac{AG}-derived asset list with the full topology assets. 
For every attack path $p=\langle h_0,\dots ,h_n\rangle$, we traverse consecutive node pairs $(h_i,h_{i+1})$. 
If their link is absent from the \ac{AG} asset list, the missing device(s) from the original topology are added to the twin assets. 
The updated asset list is then forwarded to the Cyber Twin Generator component.

\subsubsection{Cyber Twin Generator}\label{subsec:CTGenerator}
The component's inputs are the \ac{AG} and the list of facts representing the network topology (hosts, software, and connectivity) received from the AG Component Modeler component, and the catalog of GNS3 node templates available on the local hypervisor.

To generate a cyber twin, we should first select the right components.
A single breadth‑first sweep marks every node and edge on a path from an initial‑access node to an objective node. 
For each marked host, we retain its default gateway, the first‑hop DNS or Active Directory server referenced in any predicate, and the firewall enforcing its outbound rules.
The resulting subset is linear in $\mathcal{O}\bigl(|V|+|E|\bigr)$ and forms the basis of the twin.

The component then converts the subset of topology entities into a GNS3 project via raw REST calls.
It first instantiates nodes according to a deterministic lookup (e.g.,\ \texttt{\small{windows\_10}} $\rightarrow$ \texttt{\small{qemu:Win10-x64-lite}}, \texttt{\small{ios\_router}} $\rightarrow$ \texttt{\small{IOSv:15.9}}).
It then links interfaces per \texttt{\small{dataFlow}} predicate.
Finally, it uploads per-node startup scripts, enabling \emph{only} the services present in the list of topology facts.

Automated deployment assumes that every device type referenced in the lookup table is registered as a GNS3 node template on the local hypervisor. 
SCyTAG ships an importable bundle of QEMU, Docker, and IOSv templates.
These must be loaded once via the GNS3 \texttt{\small /templates} API before the first run. 
At runtime, the generator validates the presence of the template and aborts with a descriptive error if any mapping is unresolved, ensuring repeatable, license-free instantiation across hosts.  
This streamlined deployment typically cuts CPU and RAM usage by an order of magnitude (see~\cref{subsebsec:utility}), enabling generation of the cyber twin in minutes instead of hours.

The next step is host and service configuration.
Each \ac{VM} mirrors its real-world counterpart by:
\begin{itemize}[noitemsep]
    \item Installing the precise software versions associated with CVEs referenced in the \ac{AG} (downloaded at first boot via cloud-init, where feasible)
    \item Enabling just the network services flagged as relevant
    \item Reproducing system-level hardening or misconfiguration settings from the original organization’s \ac{IT} specifications 
\end{itemize}

Automating these steps eliminates manual overhead and prevents configuration drift between iterations.

The component then sets up the firewall and connectivity.
Device startup configurations enforce the minimal \acp{ACL}, \acp{VLAN}, and routing rules required for the \ac{AG} paths. 
External rules are omitted, reducing noise while preserving attack vectors.

To achieve a minimal viable cyber twin, the following selective-reduction rationale is applied:
\begin{itemize}
    \item \textbf{Host \& service filtering} - retain only hosts on \ac{AG} paths
    \item \textbf{Software \& vulnerability selection} - keep only software and CVEs explicitly referenced in an \ac{AG} path
    \item \textbf{Network-connectivity pruning} - remove links not traversed by any \ac{AG} path
\end{itemize}

The cyber twin includes only entities from the \ac{AG}; this ensures it is minimal, since only entities required for implementing the attack scenario are part of the \ac{AG}.
The cyber twin includes all the entities from the \ac{AG}; this ensures its viability, since the \ac{AG} includes all the entities required for implementing the attack scenario.

The last step is sanity validation.
Before hand-off, the component boots the twin, pings or opens a TCP socket for every \texttt{\small{dataFlow}} edge, and executes an attack to achieve a goal predicate, which represents the attacker’s objective (e.g., attackGoal(localAccess(dvr, attacker))).
This ensures that at least one attack goal is achievable.  
Sanity validation failure aborts the process and raises an alert.

The output of this module is a cyber twin environment with a minimal set of network components for attack emulation.

\subsection{Attack Emulator} \label{subsec:AttackEmulator}
The Attack Emulator module executes the scenario-specific attack on the instantiated cyber twin and provides a report for evaluation.
By binding \ac{AG} facts to Ability variables, running the attack scenario, and providing the attack results, the Attack Emulator module completes the SCyTAG workflow.

\subsubsection{Caldera Ability Variables Mapper}\label{subec:CAVM}
This component's inputs are the Caldera Abilities from the Attack Scenario Modeler and the cyber twin environment from the Cyber Twin module.
The cyber twin is represented by an Adversary YAML file with unresolved placeholders, a JSON map linking \ac{AG} node IDs to GNS3 node UUIDs, and \ac{AG} facts (e.g., host IPs and credentials).

The component assumes that all required Abilities exist, so if an Ability is missing, the process aborts with an error.
The component replaces every placeholder of the form $\texttt{\small{\#\{namespace.key\}}}$, e.g., $\texttt{\small{\#\{host.ip\}}}$ or $\texttt{\small{\#\{cred.username\}}}$, with its runtime value extracted from the \ac{AG} facts.
After substitution, each Ability is revalidated against the Caldera database. 

For every host, the component checks that the Ability’s declared \texttt{\small{platform}} (e.g., \texttt{\small{windows}}, \texttt{\small{linux}}) and \texttt{\small{executor}} fields match the host template in the cyber twin.
If there are mismatches, they are reported, and the step is skipped.

The output of this component is the set of Abilities, where the variables are filled with values.

\subsubsection{Caldera Attack Emulator} \label{subsec:CalderaAttackEmulator}
The inputs to this component are the Caldera Abilities and the cyber twin environment received from the Caldera Ability Variables Mapper module.
The component uploads the finalized YAML profile and creates a Caldera Operation (the process of running an Adversary) via the REST endpoint \texttt{\small{/api/v2/operations}}.  
The default settings are applied (0 second base delay, 2–8 seconds jitter).
These parameters can be adjusted for optimization and parallel execution.

The Ability execution steps follow the order of the \ac{AG} steps.
Branches targeting disjoint hosts are executed in parallel threads.
Shared resource conflicts are returned to a sequential order.
The emulator retries transient network errors up to three times before marking a step as failed.
Possible failure errors are:
\begin{itemize}[noitemsep]
    \item \texttt{\small NetworkFail}: Agent unreachable, retry
    \item \texttt{\small ExploitFail}: Ability returns non‑zero step fail, path may continue via alternative branch
    \item \texttt{\small DetectionFail}: Defensive block detected, remaining steps on that host are skipped, event is logged
\end{itemize}

An Operation is deemed successful if at least one objective node of the \ac{AG} is reached and all of the prerequisite steps on that path succeed.  
The success for each step is recorded individually to support a fine-grained evaluation of the proposed framework (see~\cref{subsec:EvaluationMetrics}).
Real‑time status events are streamed over Caldera’s WebSocket API.
The component's outputs are Caldera debrief reports (HTML \& JSON), and optional PCAP capture (when the \texttt{\small{--capture}} flag is set).

\subsection{Use Cases} \label{subsec:UseCases}
A lightweight, attack‑specific cyber twin enables several operational workflows beyond attack emulation.  
The SCyTAG framework can be employed in several use cases:

\begin{enumerate}[label=\textbf{\arabic*)}]
    \item \textbf{Blue‑team strengthening with MITRE D3fend:}  
    The Caldera debrief lists every successful Technique executed using the twin.  
    A postprocessor can map each Technique to its corresponding D3fend (\cite{d3fend}) countermeasure ID, highlight hosts lacking controls, and output a gap‑analysis report.
    For this use case, to get a complete gap‑analysis report and countermeasures, the Attack Path Reducer may be skipped.
        
    \textit{Outcome:} defenders know exactly which log source, block rule, or deception control to deploy and can run the scenario again to verify coverage.

    \item \textbf{SOC playbook validation:} \Acp{SOC} can load their detection and response playbooks into the twin, trigger the same \ac{CTI}‑derived attack, and measure mean time to detect, contain, and recover.  
    Because the framework's cyber twin resets in minutes, the SOC can iterate until playbooks meet a target \ac{MTTR}.  
        
    \textit{Outcome:} quantified readiness metrics and a repeatable regression test for mitigation methods.

    \item \textbf{Supply‑chain vendor audits:} A network topology provided by a third‑party vendor can be imported into SCyTAG to emulate a given attack scenario and generate a remediation list.
    Vendors receive a concrete mapping from the exposed TTPs to the required fixes (patch, config, control), while the organization validates improvements by re‑running the framework's cyber twin.  
  
    \textit{Outcome:} objective, reproducible evidence of a vendor’s security posture.

    \item \textbf{Testing what-if scenarios:} When there are a few options for network topology changes, e.g., adding new hosts or changing the connections of switches and routers, the planned topologies can be tested against given attack scenarios to examine the resilience of each topology.  
  
    \textit{Outcome:} a decision-support framework for network administrators.
\end{enumerate}

\section{Evaluation} \label{sec:Evaluation}
\subsection{Evaluation Method and Metrics} \label{subsec:EvaluationMetrics}
To validate the correctness and applicability of the SCyTAG framework, we apply its entire process to two networks. 
Each network is independently processed using its respective organizational \ac{IT} specification and simulated attack scenarios derived from CTI-aligned narratives, with Techniques that apply to the network and generate an AG with at least one attack path. 
The following sections outline, step by step, how SCyTAG transforms these artifacts into a minimal viable cyber twin, capable of emulating the attack scenario.

Each network is represented by a topology and \ac{IT} specifications.
\Ac{IT} specifications include router and firewall configuration files (OpenWrt standard), and software inventory per device, either simulated or scraped from \acs{VM} or system images.
All these artifacts serve as the raw input to the SCyTAG framework, which transforms them into logical facts.
Fact generation is automated by scripts that parse the input file and create relevant facts (as mentioned in \cref{subsec:NetworkModeleBuilder}).

The inputs to the \textit{Attack Scenario Modeler} are the relevant attack scenarios to emulate.
SCyTAG extracts ATT\&CK Techniques from the given scenarios and maps the Techniques to existing \acp{IR} (for unmodeled Techniques, we generate new \acp{IR}), and Caldera Abilities.
We verify that the \acp{IR} cover all of the attack paths described in the scenario.
The Techniques are mapped to Caldera Abilities via a local dictionary, as shown in ~\cref{lst:T1548class,lst:ability-t1548}.
In cases in which no exact match is available, SCyTAG generates stub Abilities and flags them for analyst review.

\lstdefinestyle{narrow}{
  basicstyle=\ttfamily\footnotesize,  
  breaklines=true,
  breakindent=0pt,                    
  columns=fullflexible,
  keepspaces=true,
  frame=single,
  xleftmargin=1.5em,
}

\lstdefinestyle{yaml}{
  style=narrow,                       
  string=[s]{'}{'}, stringstyle=\color{blue},
  morecomment=[l]{:}, stringstyle=\color{blue}, 
  morecomment=[l]{:}, commentstyle=\color{black},
  morecomment=[l]{-},
}

\begin{lstlisting}[basicstyle=\ttfamily\scriptsize, language=Python, label=lst:T1548class, frame=single, caption=Class skeleton for T1548]
class T1548(Technique):
    def __init__(self):
        super().__init__()
        self.rulesets.append(rs.abuse_local_access)
\end{lstlisting}

\begin{lstlisting}[basicstyle=\ttfamily\tiny, language=Python, label=lst:ability-t1548, frame=single, caption=Ability metadata for T1548.002]
name: Full Bank Abuse Elevation of Privilege (Bypass UAC Medium)
ability_number: e3db134c-4aed-4c5a-9607-c50183c9ef9e
technique: T1548.002
technique_name: Abuse Elevation of Privilege
description: Abuse Elevation of Privilege to gain higher privileges.
type: manual
tags:
  - privilege_escalation
  - abuse_elevation_of_privilege
\end{lstlisting}

The \textit{Missing Facts Completer} component described in Section~\ref{subsec:MissingFactCompletor} is used to insert each fact that cannot be part of the network model.
Facts that are conditions of the IRs representing the attack scenario and cannot be part of the organization’s IT specifications will cause immediate failure in the AG generation process.
By reviewing the IRs received from the IR Modeler and the facts received from the Network Modeler, we complete the facts that are not part of Listing 1, enabling the network model facts to affect the AG generation for the given scenario.
The complete facts and \acp{IR} are fed to MulVAL to generate an \ac{AG} for each topology.
The \ac{AG} is used by the \textit{Cyber Twin Generator} component to build the GNS3 topology (as described in \cref{subsec:CTGenerator}).
Once the cyber twin is generated, the Caldera Abilities are mapped following the process described in \ref{subsec:AttackEmulator}.
The \textit{Caldera Attack Emulator} component uploads the finalized YAML file, creates the Caldera Operation, and starts the Caldera server.
To run the attack scenario, the \textit{Caldera Attack Emulator} utilizes the GNS3 API to deploy Caldera agents on specific GNS3 network nodes (those that initiate attack steps in the \ac{AG}).
Once the agents connect to the Caldera server, the Caldera Operation representing the attack scenario is launched (following the process explained in \cref{subsec:CalderaAttackEmulator}).

We evaluate \textsc{SCyTAG} using the metrics described below, which are applied to the two topologies, in both \emph{full} and \emph{reduced} form.

\subsubsection{Utility - Reduction and Efficiency} \label{subsebsec:utility}
A scalable, attack‑specific cyber twin must (i) reduce the resources needed to represent the network under test, and (ii) retain the entities required for attack emulation.
Utility metrics quantify the first goal's structural and runtime savings and are measured directly from topology files and virtualization logs.
The second goal is evaluated with the help of the \ac{AG}.

Table~\ref{tab:utility-metrics} summarizes the concrete structural and runtime metrics we collect.
The metrics presented in this table demonstrate how the cyber twin reduces infrastructure costs without compromising security-relevant behavior.

\newcolumntype{L}[1]{>{\raggedright\arraybackslash}p{#1}}
\newcolumntype{C}[1]{>{\centering\arraybackslash}p{#1}}

\begin{table*}[htbp]
  \centering
  \caption{Utility metrics and collection procedures.}
  \label{tab:utility-metrics}
  \renewcommand{\arraystretch}{1.05}
  \setlength{\tabcolsep}{3pt}
  \begin{tabularx}{\textwidth}{@{}L{0.09\linewidth}L{0.14\linewidth}%
    C{0.19\linewidth}L{0.27\linewidth}L{0.26\linewidth}@{}}
    \toprule
    \textbf{Category} & \textbf{Metric} & \textbf{Formula / Definition} & \textbf{What it means} &
    \textbf{Collection pipeline} \\ \midrule
    \multirow{2}{*}{Structural} 
        & Node ratio & $\bigl(|H_{\mathrm{full}}|-|H_{\mathrm{red}}|\bigr)/|H_{\mathrm{full}}|$ 
        & Percentage of hosts eliminated by reduction 
        & Static count from twin project files \\
        & Edge ratio & $\bigl(|E_{\mathrm{full}}|-|E_{\mathrm{red}}|\bigr)/|E_{\mathrm{full}}|$ 
        & Percentage of L2/L3 links eliminated & idem \\ \addlinespace
    \multirow{5}{*}{Runtime} 
        & Avg CPU~$\Delta$ & $\bigl(\overline{\mathrm{CPU}}_{\mathrm{full}}-\overline{\mathrm{CPU}}_{\mathrm{red}}\bigr)\!/\overline{\mathrm{CPU}}_{\mathrm{full}}$
        & Mean CPU utilized on the hypervisor & \texttt{vmstat}/\texttt{pidstat} (1 s) \\[3pt]
        & Peak RAM~$\Delta$ & relative drop in maximum resident set size (RSS) & Memory head room gained & \texttt{smem} (5 s) \\[3pt]
        & Disk I/O~$\Delta$ & relative drop in bytes\textsubscript{read+write} & Storage traffic saved & \texttt{iostat k 1} (QCOW2) \\[3pt]
        & Boot time~$\Delta$ & $t_{\!ready}^{\mathrm{full}}-t_{\!ready}^{\mathrm{red}}$ & Time saved between runs & GNS3 REST events \\[3pt]
        & Runtime accel. ($R_T$) 
          & $\mathrm{Duration}_{\mathrm{full}}/\mathrm{Duration}_{\mathrm{red}}$ 
          & Acceleration factor of the entire operation &
          Unix \texttt{date} stamps \texttt{start}/\texttt{finish} \\ 
    Energy proxy & $E$ consumption~$\Delta$ 
        & $\bigl(\overline{C}_{\mathrm{full}}-\overline{C}_{\mathrm{red}}\bigr)\,P_{\mathrm{vCPU}}\;t$  
        & Watt hour saving (guest CPU proxy) 
        & Azure “Guest CPU\%” metric+runtime $t$ \\ \bottomrule
    \bottomrule
  \end{tabularx}
\end{table*}

\paragraph*{Structural utility}
This metric assesses whether SCyTAG has successfully reduced the topology.

\paragraph*{Runtime and energy}  
Let $\bar x_{\text{full}}$ and $\bar x_{\text{twin}}$ be the \emph{mean} of a resource utilization trace (CPU, RAM, I/O), and let
$t_{\text{full}}$, $t_{\text{twin}}$ be the duration times until the
\textsc{ready} event is triggered.
For every resource, we report the \emph{relative saving}

\[
\Delta x[\%] \;=\;
\frac{\bar x_{\text{full}} - \bar x_{\text{twin}}}{\bar x_{\text{full}}}\times 100 .
\]

\noindent Boot time acceleration is a simple difference
$\Delta t_{\text{boot}} = t_{\text{full}}-t_{\text{twin}}$.
According to Microsoft’s published TDP for a Standard~D4s~v3 \ac{VM}
($P_\mathrm{vCPU}=11.5$W), the saved energy is as follows:

\[
\Delta E = (\bar{\mathrm{CPU}}_{\text{full}}-\bar{\mathrm{CPU}}_{\text{twin}})
           \times P_\mathrm{vCPU}\times t_{\text{twin}}
\]

\subsubsection{Effectiveness - Fidelity and Success} \label{subsec:effectiveness}
Effectiveness metrics quantify how closely the cyber twin reproduces the attacker's behavior observed on the full topology.  
They are computed from paired Caldera debrief reports.
Table~\ref{tab:effectiveness-metrics} summarizes the effectiveness metrics computed from paired debrief reports.

\begin{table*}[htbp]
  \centering
  \caption{Effectiveness metrics extracted from paired debrief reports.}
  \label{tab:effectiveness-metrics}
  \renewcommand{\arraystretch}{1.10}
  \begin{tabular}{l c p{0.33\linewidth} p{0.30\linewidth}}
    \toprule
    \textbf{Symbol} & \textbf{Metric} & \textbf{Formula / Description} & \textbf{Evaluation Objective} \\ \midrule
    ASP & Ability Success Parity & $\bigl|\mathrm{Succ}_{\mathrm{full}}\cap\mathrm{Succ}_{\mathrm{red}}\bigr| \big/ |\mathrm{Succ}_{\mathrm{full}}|$ & Confirm that every Ability that succeeds on the full network also succeeds on the twin \\[4pt]
    TCP & Technique Coverage Parity & Jaccard similarity between ATT\&CK Technique sets & Ensure the twin exposes the same adversarial technique surface as the full network \\[4pt]
    PES & Path Equivalence Score & $\dfrac{\text{\# edges reproduced}}{\text{\# edges in full AG}}$ & Verify that structural attack paths present in the full \ac{AG} are retained in the twin \\[4pt]
    $\tau$ & Order Similarity & Normalized Kendall-$\tau$ between ability sequences & Check that the causal execution order of abilities is preserved between runs \\[4pt]
    $\Delta\text{Obj}$ & Objective Discrepancy & $0$ if goal reached in both runs, else $1$ & Validate that the attack achieves its end-to-end objective in both environments \\ \bottomrule
  \end{tabular}
\end{table*}

Debrief processing is performed by a script (\texttt{\small{compare\_debrief.py}}) that (1) loads the two JSON reports, (2) normalizes timestamps, (3) computes the metrics, and (4) generates CSV tables and optional bar charts.

To evaluate the SCyTAG framework, we use two representative network environments that vary in complexity, origin, and scale.
Each topology enables us to evaluate both the efficiency and precision of the framework under various structural and operational constraints.

These two topologies cover different use cases: one is a real enterprise network, and the other is a fictitious, complex organizational network. They allow us to evaluate the generalization and robustness of the SCyTAG framework.

\subsection{Test Environment}
The evaluation test cases were run on a cloud environment.
Its characteristics are presented in \cref{tab:test-env-characteristics}.

\begin{table}[htbp]
  \centering
  \caption{Test environment characteristics.}
  \label{tab:test-env-characteristics}
  \begin{tabular}{c c c c c}
    \toprule
    No. vCPUs & RAM & Disk & VM Architecture & OS \\
    \midrule
    4 & 16GB & 128GB & x64 & Ubuntu 24.04 \\
    \bottomrule
  \end{tabular}
\end{table}

\subsection{Case Study 1 - UK Office Topology}
The UK Office topology is a real enterprise network.
It reflects a mid-scale organizational setup with standard departmental segmentation (e.g., Admin, Core, Storage).

\begin{figure}[htbp]
\centering
\includegraphics[width=0.85\linewidth]{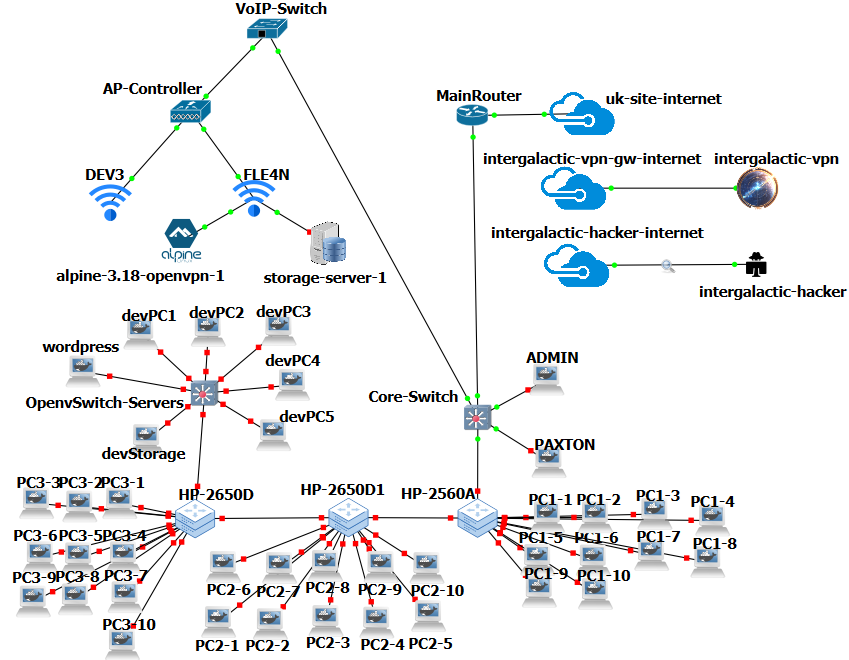}
\caption{Full UK Office topology (GNS3-based).}
\label{fig:UkOffice-full}
\end{figure}

\begin{figure}[htbp]
\centering
\includegraphics[width=0.85\linewidth]{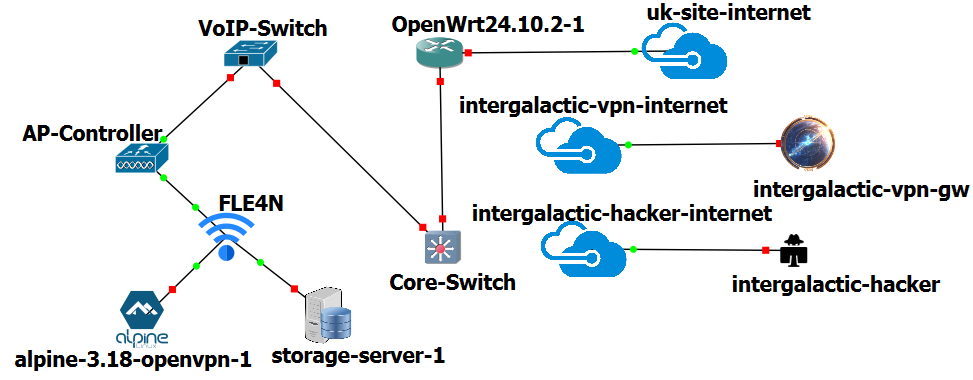}
\caption{Reduced UK Office topology (GNS3-based).}
\label{fig:UkOffice-reduced}
\end{figure}

As seen in \cref{fig:UkOffice-full}, the full topology contains 54 nodes and 53 edges. 
The reduced form is shown in \cref{fig:UkOffice-reduced}, and it contains 10 nodes and 9 edges.
This represents a reduction ratio of $\approx$81.5\% in nodes and $\approx$83\% in edges.
The reduced topology preserves only those nodes and connections relevant to the \ac{AG} derived for the given \ac{CTI} scenario. 
It includes all intermediary nodes on viable attack paths, as well as supporting infrastructure such as firewalls and domain controllers.

\vspace{-4mm}

\vspace{-4mm}
\subsubsection{Network Modeler}
All topology artifacts are transformed into logical facts.
The topology is modeled using 146 facts, which utilize the eight fact types listed in \cref{lst:BasicFacts}.

\subsubsection{Attack Scenario Modeler}
The attack scenario related to the UK Office topology is modeled with 21 \acp{IR} corresponding to certificate forgery, web UI exploitation, and VPN-based lateral movement.

\subsubsection{Attack Graph Generator}
The raw fact base comprises 146 facts. 
The completer adds 112 facts (77\%) to satisfy \ac{IR} preconditions that could not be satisfied by the given fact types.

\begin{figure*}[htbp]
    \centering
    \begin{subfigure}[b]{0.4\textwidth}
    \centering
    \includegraphics[width=\columnwidth]{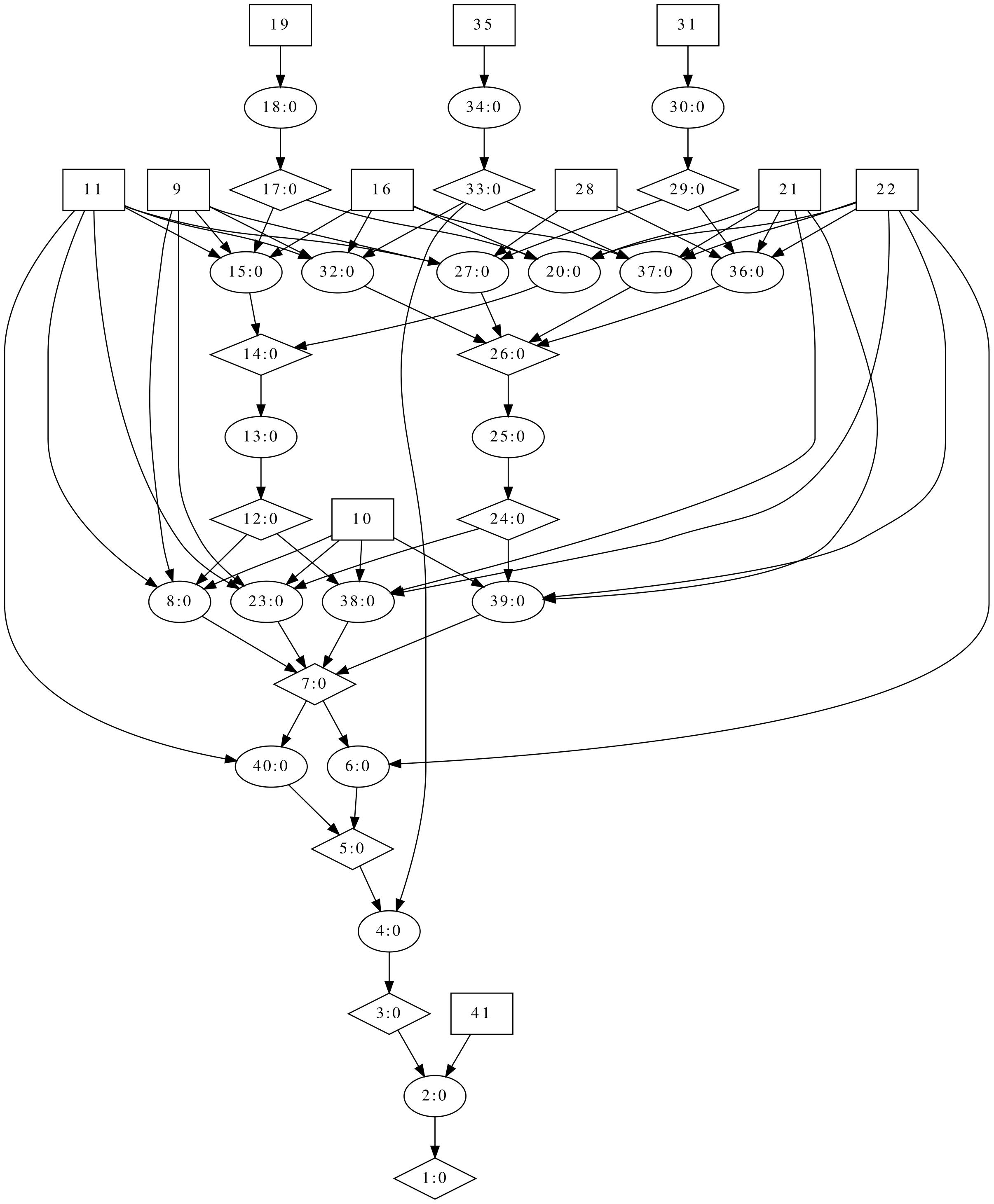}
    \caption{\ac{AG} structure.}
    \label{fig:UkOffice-graph}
  \end{subfigure}
  \hfill
  \begin{subfigure}[b]{0.59\textwidth}
    \centering
    \begin{adjustbox}{max width=\linewidth}
    \begin{lstlisting}[basicstyle=\ttfamily\scriptsize, frame=none]
    <1>:lateralMovementVPN('intergalactic-hacker','intergalactic-vpn-gw','alpine-openvpn-1')
    (2):RULE 22 (Lateral movement over VPN network.)
    [3]:canAccessVPN('intergalactic-hacker','intergalactic-vpn-gw')
    (4):RULE 21 (VPN network access.)
    [5]:canCreateValidVPNCertificate('intergalactic-hacker','intergalactic-vpn-gw')
    (6),(40):RULE 20 (Forge VPN certificate.)
    [7]:softwareCompromisedLocally('intergalactic-vpn-gw')
    (8),(23),(38),(39):RULE 19 (Local privilege escalation.)
    [9]:vulExists('cve-2023-27524','intergalactic-web-ui','2.4.12',network,...)
    [10]:hasAccount(root,'intergalactic-vpn-gw','operating-system-administration-account')
    [11]:residesOn('intergalactic-vpn-gw','intergalactic-web-ui','2.4.12')
    <12>:compromisedVPNClient('alpine-openvpn-1','intergalactic-vpn-gw')
    (13),(25):RULE 15 (Host compromised.)
    [14]:softwareCompromisedRemotely('alpine-openvpn-1','intergalactic-vpn-gw')
    (15),(20),(27):RULE 18 (Remote software compromise.)
    [16]:networkService('intergalactic-vpn-gw','intergalactic-web-ui',udp,1194,root)
    [17]:netAccess(alice,'alpine-openvpn-1','intergalactic-vpn-gw',udp,1194)
    (18),(30),(34):RULE 5 (Net direct access.)
    [19]:hasAccess(alice,'alpine-openvpn-1','intergalactic-vpn-gw',udp,1194)
    [21]:vulExists('cve-zero-day-web-ui-execute','intergalactic-web-ui','0.1.0rc0',...)
    [22]:residesOn('intergalactic-vpn-gw','intergalactic-web-ui','0.1.0rc0')
    <24>:compromisedVPNClient('intergalactic-hacker','intergalactic-vpn-gw')
    [26]:softwareCompromisedRemotely('intergalactic-hacker','intergalactic-vpn-gw')
    [28]:networkService('intergalactic-vpn-gw','intergalactic-web-ui',http,80,root)
    [29]:netAccess(hacker,'intergalactic-hacker','intergalactic-vpn-gw',http,80)
    [31]:hasAccess(hacker,'intergalactic-hacker','intergalactic-vpn-gw',http,80)
    (32),(36),(37):RULE 18 (Remote software compromise.)
    [33]:netAccess(hacker,'intergalactic-hacker','intergalactic-vpn-gw',udp,1194)
    [35]:hasAccess(hacker,'intergalactic-hacker','intergalactic-vpn-gw',udp,1194)
    [41]:inSubnet('alpine-openvpn-1',vpn)
    \end{lstlisting}
    \end{adjustbox}
    \caption{Node interpretation.}
    \label{fig:UkOffice-list}
    \end{subfigure}
  \caption{Attack graph generated for the UK Office topology.}
  \label{fig:UkOffice-ag}
\end{figure*}

The complete set of facts and \acp{IR} are fed to MulVAL to generate an \ac{AG}.
As visualized in \cref{fig:UkOffice-ag}, the attack begins with the attacker forging a client certificate and tunneling into the \texttt{\small intergalactic-vpn-gw} (T1190).  
Two web-UI CVEs (e.g., CVE-2023-27524) enable privilege escalation on the gateway. 
The attacker then pivots laterally to \texttt{\small alpine-openvpn-1} and adjacent subnets, completing five-hop path to root (final goal).  
The \ac{AG} contains 41 nodes, each of which is part of at least one scenario step.

\lstdefinestyle{aglist}{
  basicstyle=\ttfamily\scriptsize,
  breaklines=true,
  breakatwhitespace=false,   
  columns=fullflexible,      
  keepspaces=true,
  breakindent=0pt,
  postbreak=\mbox{\tiny$\hookrightarrow$}, 
  frame=lines,
  numbers=none
}

\subsubsection{Cyber Twin}
Every host in the cyber twin runs a dedicated Caldera agent that inherits the host’s native subnet address (internet or intranet subnets).
For example, on the UK Office topology, the Caldera agent that reflects the attacker device resides in an 'internet' environment, which is located outside the UK Office topology network and provides ingress connectivity to the network.


The reduced UK Office topology consists of 10 nodes: 4 Ubuntu Docker container hosts, 1 OpenWrt firewall, 1 OpenWrt Layer 3 switch, 3 components of Wi-Fi connectivity representation (using the VoIP-Switch, AP-Controller, and FLE4N), and the Cloud node, which represents the egress connectivity and internet connectivity.
The network contains two \acp{VLAN}.

\subsubsection{Attack Emulator}
Each host in the cyber twin starts a \textit{Sandcat} agent inside its original \ac{VLAN} segment as dictated by the \ac{AG} (e.g., the VPN-GW agent in the VPN subnet, \textit{alpine-openvpn-1} in the IoT subnet).
The agent beacons to the Caldera server over the same routed path that exists in the production network.
The ICMP and TCP/8888 handshakes succeed within 30 seconds.  
The adversary profile (\texttt{\small{ukoffice\_op.yml}}) is imported without missing variables, and the smoke test passes on the agent (inter-connectivity).

Caldera then replays the eight-step adversary profile (\texttt{\small ukoffice\_op.yml}):

\begin{enumerate}[label=\arabic*) ,nosep,leftmargin=*]
    \item web-UI version discovery (\textbf{T1016})  
    \item remote shell injection via web-UI flaw (\textbf{T1505.003})
    \item credential dump from the VPN database (\textbf{T1555})
    \item password cracking with John (\textbf{T1110.002})
    \item client certificate forgery \& VPN connect (\textbf{T1133})
    \item VPN network scan (\textbf{T1423})
    \item SSH pivot to an internal host (\textbf{T1021.004})
    \item SSH tunnel into the VLAN core (\textbf{T1021.004})
\end{enumerate}

For step-level parity, we run the corresponding Abilities on the associated scenario.
The full and reduced UK Office topologies execute the same \textbf{8} Abilities, all with exit code~0, yielding \ac{ASP}=1.00 and \ac{TCP}=1.00. 

The identical results presented in \Cref{tab:UkOffice-ability} confirm that SCyTAG’s reduction removes only irrelevant infrastructure while preserving every action the adversary performs on the full network.

\begin{table}[htbp]
\scriptsize
\caption{Key Abilities executed in the full network vs.\ twin (UK Office topology).}
\label{tab:UkOffice-ability}
\centering
\begin{tabular}{@{}l l l c c@{}}
\toprule
Scenario step & Technique & Host & Full & Twin \\
\midrule
Web-UI version probe          & T1016      & intergalactic-vpn & \checkmark & \checkmark \\
Web-UI RCE (shell upload)     & T1505.003  & intergalactic-vpn & \checkmark & \checkmark \\
Credential dump (VPN DB)      & T1555      & intergalactic-vpn & \checkmark & \checkmark \\
Password cracking (John)      & T1110.002  & analyst \ac{VM}          & \checkmark & \checkmark \\
Client-certificate forgery    & T1133      & attacker workstat. & \checkmark & \checkmark \\
VPN network scan              & T1423      & 10.8.0.0/24         & \checkmark & \checkmark \\
SSH pivot to internal host    & T1021.004  & 10.8.10.2           & \checkmark & \checkmark \\
SSH tunnel into VLAN core     & T1021.004  & CoreSwitch subnet   & \checkmark & \checkmark \\
\bottomrule
\end{tabular}
\end{table}

\subsubsection{Utility - Reduction and Efficiency}

\paragraph*{Structural utility}
The results for this metric, which quantifies the extent to which the cyber twin is able to reduce its source topology without affecting the attacker’s shortest route to the goal, are presented in Table~\ref{tab:structural-reduction-UkOffice}.
As can be seen, SCyTAG successfully reduced the topology, decreasing both the number of hosts and connections.

\begin{table}[htbp]
  \centering
  \caption{Structural reduction achieved by SCyTAG (UK Office topology).}
  \label{tab:structural-reduction-UkOffice}
  \renewcommand{\arraystretch}{1.15}
  \begin{tabular}{@{}l rrr rrr@{}}
    \toprule
    & \multicolumn{3}{c}{\textbf{Hosts}} & \multicolumn{3}{c}{\textbf{Connections}} \\
    \cmidrule(lr){2-4} \cmidrule(l){5-7}
    Topology & Full & Twin & $R_H\%$ & Full & Twin & $R_E\%$ \\
    \midrule
    UK Office & 54 & 10 & 81.48 & 53 & 9 & 83.02 \\
    \bottomrule
  \end{tabular}
\end{table}

In the UK Office cyber twin, the critical path retains the same hop length as the full graph (five hops). 
Ten of the 54 nodes and 9 of the 53 connections are kept, with reduction ratios of $R_H=81.48\%$ and $R_E=83.02\%$, respectively.

\paragraph*{Runtime and energy}  
For the UK Office topology, according to 'full\_cpu.log' and 'reduced\_cpu.log', $\bar{\mathrm{CPU}}_{\text{full}}=13.64\%$,  $\bar{\mathrm{CPU}}_{\text{reduced}}=11.11\%$, therefore $\Delta\text{CPU}=18.55\%$.
Similarly, $\bar{\mathrm{RAM}}_{\text{full}}=0.134$GB and $\bar{\mathrm{RAM}}_{\text{reduced}}=0.129$GB, yielding $\Delta\text{RAM}=3.71\%$, where as the read / write counters are identical, so $\Delta\text{I/O}=0\%$.
We measure the duration of the Caldera Operation: $t_{\text{full}}=736$ seconds and $t_{\text{reduced}}=456$ seconds, presenting $\Delta t_{\text{boot}} = 280$ seconds.
Finally, \[
E = \frac{\bar{C}}{100}\;P_{\text{vCPU}}\;\frac{t}{3600}\;\text{Wh},
\]
so for the UK Office runs
\[
\begin{aligned}
E_{\text{full}} &= 0.1364 \times 11.5 \times \frac{736}{3600} \;=\; 0.3207~\text{Wh},\\[2pt]
E_{\text{twin}} &= 0.1111 \times 11.5 \times \frac{456}{3600} \;=\; 0.1602~\text{Wh},\\[2pt]
\Delta E        &= 0.3207 - 0.1602 = 0.1605~\text{Wh}\;(\textbf{-49.95\%}).
\end{aligned}
\]

The cyber twin, therefore, cuts the guest CPU energy by almost one-half while maintaining all attacker functionality.
The Delta CPU, RAM, I/O, runtime, and energy (between the full and reduced topologies) are presented in Table~\ref{tab:runtime-deltas-UK-Office}.

\begin{table}[htbp]
    \caption{Runtime and energy savings (UK Office topology).}
    \label{tab:runtime-deltas-UK-Office}
    \small
    \setlength{\tabcolsep}{2pt}
    \renewcommand{\arraystretch}{1.15}
    \begin{tabular*}{\linewidth}{@{\extracolsep{\fill}}%
                              l r r r r r @{}}
        \toprule
        Topology &
        $\Delta$CPU\% &
        $\Delta$RAM\% &
        $\Delta$I/O\% &
        \shortstack{$\Delta t$ (sec)} &
        \shortstack{$\Delta E$ (\%)} \\
        \midrule
        UK Office & 18.55 & 3.73 & 0 & 280 & 49.95 \\
        \bottomrule
        \end{tabular*}
\end{table}

\subsubsection{Effectiveness - Fidelity and Success}
The effectiveness metrics confirm that the UK Office twin executes \emph{all} critical Abilities present in the full run.
For this topology, ASP=1.00, and it preserves 100\% of the Techniques (TCP=1.00).  
The \ac{PES}=1.00, and the order similarity is \(\tau=1.00\). 
The results indicate that reduction neither deletes nor reorders causal steps.  
The final objective is reached (\(\Delta\mathrm{Obj}=0\)).

These invariants show that SCyTAG prunes only the topology entities that are not affected by the attack scenario, preserving attack behavioral fidelity while reducing the attack emulation cost.

\subsection{Case Study 2 - Bank Topology}
The Bank topology is a synthetic topology modeled in GNS3 to simulate the internal architecture of a large-scale financial organization. 
This environment was specifically designed to examine SCyTAG's scalability and precision on complex deployments with numerous endpoint classes (e.g., HR, Marketing, and Accounting \acp{VLAN}).
We evaluate the framework's ability to operate across deep service stacks, multiple routers, and diverse access control structures.

\begin{figure}[htbp]
\centering
\includegraphics[width=0.99\linewidth]{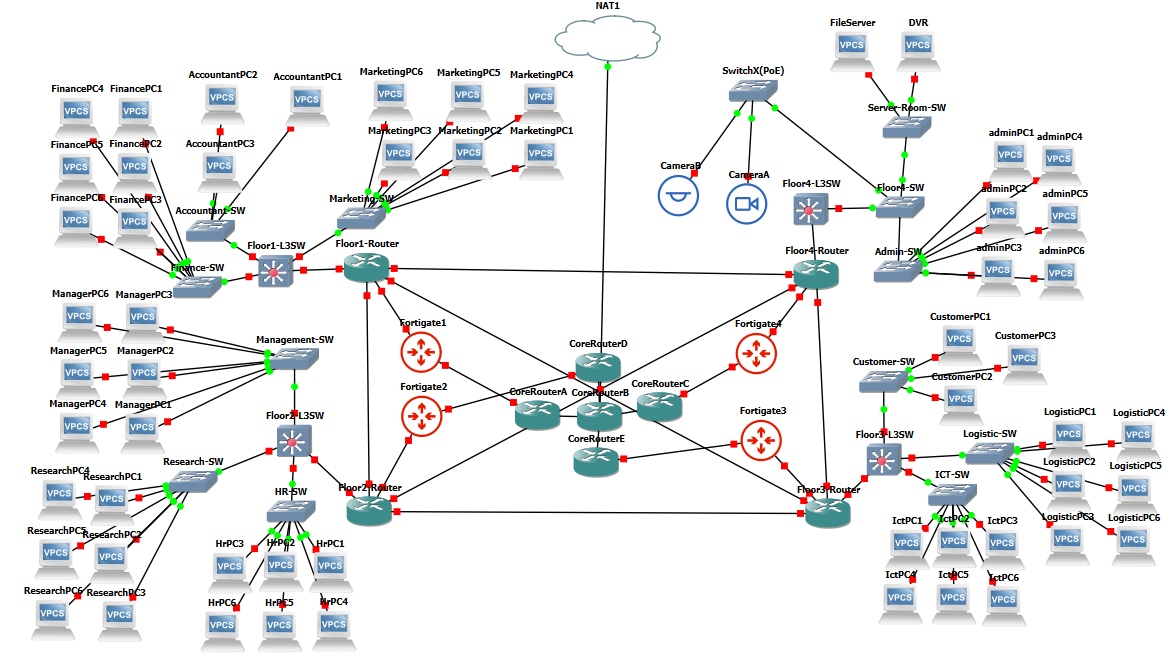}
\caption{Full Bank topology (GNS3-based).}
\label{fig:bank-full}
\end{figure}

\begin{figure}[htbp]
\centering
\includegraphics[width=0.9\linewidth]{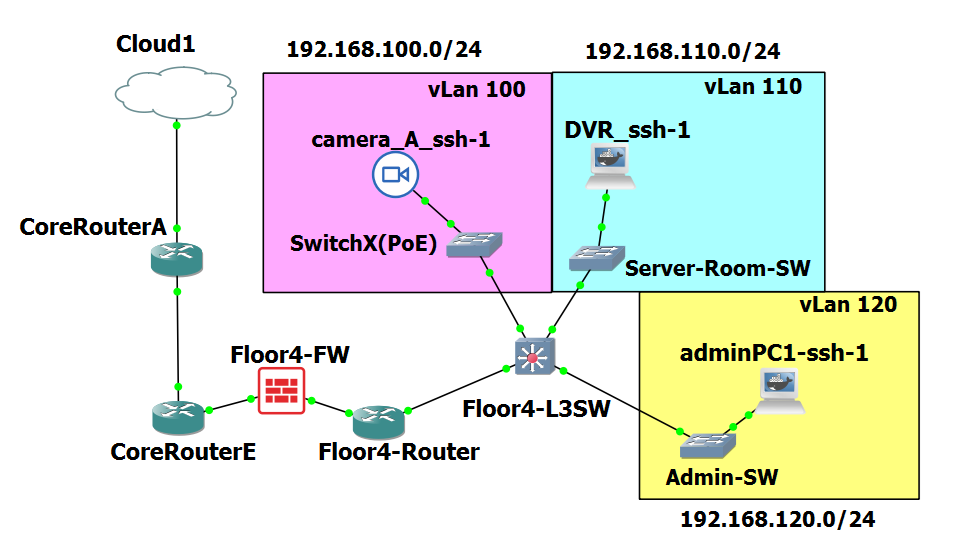}
\caption{Reduced Bank topology (GNS3-based).}
\label{fig:bank-reduced}
\end{figure}

As seen in \Cref{fig:bank-full}, the full topology contains 95 nodes and 94 edges.  
The reduced form is shown in \cref{fig:bank-reduced}, and it contains 12 nodes and 11 edges,\footnote{Twenty-four nodes appear on at least one logical \ac{AG} edge. Consolidation of infrastructure roles reduces the twin's size to 15 \acp{VM} without loss of fidelity.}.
This represents a reduction ratio of $\approx\!88\%$.  
This reduction demonstrates SCyTAG’s capacity to distill a high-fidelity cyber twin from a highly interconnected, operationally dense infrastructure.

\subsubsection{Network Modeler}
All topology artifacts were transformed into logical facts.
The topology is modeled using 164 facts, which utilize the eight fact types listed in \cref{lst:BasicFacts}.

\subsubsection{Attack Scenario Modeler}
The attack scenario related to the Bank topology is modeled with 11 \acp{IR} corresponding to credential file theft, ingress tool transfer, DNS-based \ac{MITM} spoofing, enabled delegated code execution, and root shell on the \ac{DVR}.

\subsubsection{Attack Graph Generator}
The raw fact base comprises 164 facts. 
The completer adds 43 facts (26.2\%) to satisfy \ac{IR} preconditions that could not be satisfied by the given fact types.

\begin{figure*}[htbp]
    \centering
    \begin{subfigure}[b]{0.59\textwidth}
    \centering
    \includegraphics[width=\columnwidth]{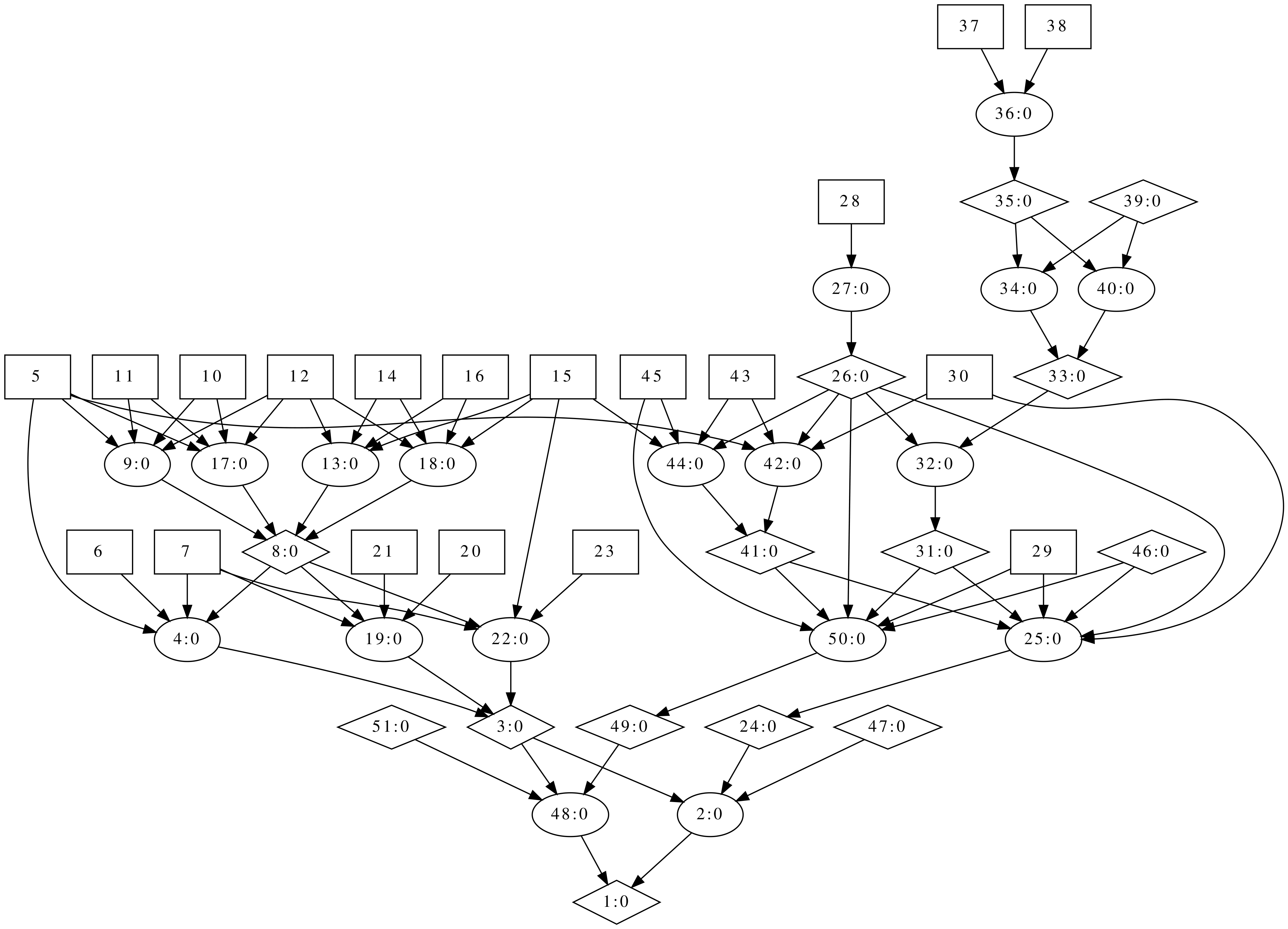}
    \caption{The \ac{AG} structure.}
    \label{fig:bank-ag-A}
    \end{subfigure}
    \centering
    \begin{subfigure}[b]{0.35\textwidth}
    \centering
    \begin{lstlisting}[basicstyle=\ttfamily\tiny, frame=none]
    <1>:fullCampaign(attacker,adminPC1,cameraA,'DVR')
    (2),(48):RULE 25 (Exploit EternalBlue from cameraA to DVR)
    <3>:execDelegatedCode(attacker,cameraA,'DVR',root)
    (4),(19),(22):RULE 24 (Exploit EternalBlue from cameraA to DVR)
    [5]:vulExists(arpSpoofVuln,arpd,ver1,network,caLoss,critical)
    [6]:networkService('DVR',arpd,tcp,'445',root)
    [7]:hasAccess(attacker,cameraA,'DVR',tcp,'445')
    <8>:compromised(cameraA)
    (9),(13),(17),(18):RULE 16 (compromised host 'cameraA')
    [10]:maliciousInteraction(cameraA,attacker,arpd)
    [11]:residesOn(cameraA,arpd,ver1)
    [12]:deviceOnline(cameraA,'Linux')
    [14]:maliciousInteraction(cameraA,attacker,sshd)
    [15]:vulExists(sshdSpoofVuln,sshd,ver1,network,caLoss,critical)
    [16]:residesOn(cameraA,sshd,ver1)
    [20]:vulExists(cve_2017_0144,smbV1,ver1,network,caLoss,critical)
    [21]:networkService('DVR',smbV1,tcp,'445',root)
    [23]:networkService('DVR',sshd,tcp,'445',root)
    <24>:ingressToolTransfer(arpd,attacker,cameraA,_,'22')
    (25),(50):RULE 17 (Net access hop)
    <26>:netAccess(attacker,adminPC1,cameraA,tcp,'22')
    (27):RULE 6 (Net direct access)
    [28]:hasAccess(attacker,adminPC1,cameraA,tcp,'22')
    [29]:hacl(adminPC1,cameraA,tcp,'22')
    [30]:networkService(cameraA,arpd,tcp,'22',root)
    <31>:dataInject(attacker,cameraA,'/etc/shadow',_,'22')
    (32):RULE 19 (Can access to the local file on the host)
    <33>:accessFile(attacker,cameraA,root,read,'/etc/shadow')
    (34),(40):RULE 10 (Valid file protection mechanism)
    <35>:localFileProtection(cameraA,'/etc/shadow',root,read)
    (36):RULE 11 (Valid file protection mechanism)
    [37]:ownerAccessible(cameraA,read,'/etc/shadow')
    [38]:fileOwner(cameraA,'/etc/shadow',root)
    <39>:execCode(attacker,cameraA,root)
    <41>:localAccess(attacker,cameraA,root)
    (42),(44):RULE 9 (Privilege escalation using setuid program)
    [43]:malicious(attacker)
    [45]:networkService(cameraA,sshd,tcp,'22',root)
    <46>:mitmE2E(attacker,adminPC1,cameraA,'DVR',tcp,'22')
    <47>:credentialsAccessInFiles(arpd,adminPC1)
    <49>:ingressToolTransfer(sshd,attacker,cameraA,_,'22')
    <51>:credentialsAccessInFiles(sshd,adminPC1)
    \end{lstlisting}
    \caption{\ac{AG} node interpretation.}
    \label{fig:bank-ag-B}
    \end{subfigure}
    \caption{Attack graph generated for the Bank topology.}
    \label{fig:bank-ag}
\end{figure*}

The complete set of facts and \acp{IR} are fed to MulVAL to generate an \ac{AG}.
As shown in \cref{fig:bank-ag}, the attacker first reads \texttt{\small /etc/shadow} on \texttt{\small cameraA} (T1552.001), then mounts a payload to the \texttt{\small DVR} (T1105), and triggers remote code execution via EternalBlue (CVE-2017-0144), all within \acp{VLAN} 100 → 110 → 120.  
MulVAL instantiates 11 \acp{IR}, yielding a seven-hop critical path and an \ac{AG} of about 60 nodes.

\lstdefinestyle{aglist}{
  basicstyle=\ttfamily\scriptsize,
  breaklines=true,
  breakatwhitespace=false,   
  columns=fullflexible,      
  keepspaces=true,
  breakindent=0pt,
  postbreak=\mbox{\tiny$\hookrightarrow$}, 
  frame=lines,
  numbers=none
}

\subsubsection{Cyber Twin}
Every host in the cyber twin runs a dedicated Caldera agent that inherits the host’s native subnet address.
For example, on the Bank topology, the Caldera agent that reflects cameraA device resides in the same \ac{VLAN}, which is 192.168.100.0/24. 
Those agents reach the Caldera server on Cloud1 via the lab’s routed backbone. 

The reduced Bank topology consists of 12 nodes: 3 Ubuntu Docker container hosts, 3 IOSv routers, 1 OpenWrt firewall, 1 OpenWrt Layer 3 switch, 3 Layer 2 switches, and the Cloud node, which represents the egress connectivity.
The network contains three \acp{VLAN} (100, 110, 120).

\subsubsection{Attack Emulator}
All three Floor-4 agents register within 60 seconds, each constrained to the \ac{VLAN} where its host resides in the \ac{AG}: camera/DVR agents in \ac{VLAN} 100, server agents in \ac{VLAN} 110, and admin-PC agents in \ac{VLAN} 120.
Traffic passes through the existing router–firewall chain to the Caldera server, which sits outside the emulated network, exactly as it would in the full topology.  
The adversary profile (\texttt{\small{bank\_floor4\_op.yml}}) replays the identical Ability sequence used in the full network, imported without missing variables, and the functional smoke tests pass on the agents.

For step-level parity, we run the corresponding Abilities on the associated scenario.
The full and reduced Bank topologies execute the same \textbf{3} Abilities, all with exit code~0, yielding ASP=1.00 and TCP=1.00.
The identical results presented in \Cref{tab:bank-ability} confirm that SCyTAG’s reduction removes only irrelevant infrastructure while preserving every action the adversary performs on the full network.

\begin{table}[htbp]
    \footnotesize
    \caption{Key Abilities executed in the full network vs. twin (Bank topology).}
    \label{tab:bank-ability}
    \centering
    \begin{tabular}{@{}l l l c c@{}}
        \toprule
        Scenario step & Technique & Host & Full & Twin \\\midrule
        Retrieve Password & T1552.001 & adminPC1 & \checkmark & \checkmark \\
        Transfer File & T1105 & cameraA  & \checkmark & \checkmark \\
        Remote Shell  & T1059.004  & DVR & \checkmark & \checkmark \\ \bottomrule
    \end{tabular}
\end{table}

\subsubsection{Utility - Reduction and Efficiency}

\paragraph*{Structural utility}
SCyTAG successfully reduced the topology without affecting the attacker’s shortest route to the goal.
Table~\ref{tab:structural-reduction-bank} quantifies the extent to much the cyber twin is able to reduce its source topology.  

\begin{table}[htbp]
  \centering
  \caption{Structural reduction achieved by SCyTAG (Bank topology).}
  \label{tab:structural-reduction-bank}
  \renewcommand{\arraystretch}{1.15}
  \begin{tabular}{@{}l rrr rrr@{}}
    \toprule
    & \multicolumn{3}{c}{\textbf{Hosts}} & \multicolumn{3}{c}{\textbf{Connections}} \\
    \cmidrule(lr){2-4} \cmidrule(l){5-7}
    Topology & Full & Twin & $R_H\%$ & Full & Twin & $R_E\%$ \\
    \midrule
    Bank   & 95 & 12 & 87.37 & 96 & 11 & 88.54 \\
    \bottomrule
  \end{tabular}
\end{table}

In the Bank cyber twin, the critical path remains seven hops. 
The exact three \acp{VLAN} (100, 110, 120) that appear on at least one \ac{AG} edge are preserved. 
Twelve of the 95 nodes and 11 of the 96 connections are kept, with reduction ratios of $R_H=87.37\%$ and $R_E=88.54\%$, respectively.  

\paragraph*{Runtime and energy}  
For the Bank topology, according to 'full\_cpu.log' and 'reduced\_cpu.log', $\bar{\mathrm{CPU}}_{\text{full}}=40.77\%$,  $\bar{\mathrm{CPU}}_{\text{reduced}}=21.00\%$, hence $\Delta\text{CPU}=48.49\%$.
Similarly, $\bar{\mathrm{RAM}}_{\text{full}}=2.397$GB and $\bar{\mathrm{RAM}}_{\text{reduced}}=0.809$GB, yielding $\Delta\text{RAM}=66.25\%$, where as the read / write counters are identical, so $\Delta\text{I/O}=0\%$.
We measure the duration of the Caldera operation: $t_{\text{full}}=337$ seconds and $t_{\text{reduced}}=252$ seconds, presenting $\Delta t_{\text{boot}} = 85$ seconds.
Finally, \[
E = \frac{\bar{C}}{100}\;P_{\text{vCPU}}\;\frac{t}{3600}\;\text{Wh},
\]
so for the Bank runs
\[
\begin{aligned}
E_{\text{full}} &= 0.408 \times 11.5 \times \frac{337}{3600} \;=\; 0.439~\text{Wh},\\[2pt]
E_{\text{twin}} &= 0.210 \times 11.5 \times \frac{252}{3600} \;=\; 0.169~\text{Wh},\\[2pt]
\Delta E        &= 0.439 - 0.169 = 0.270~\text{Wh}\;(\textbf{-61.54\%}).
\end{aligned}
\]

The cyber twin, therefore, cuts guest CPU energy by almost one-half while maintaining all attacker functionality.
Delta CPU, RAM, I/O, runtime, and energy (between full and reduced topologies) are presented in Table~\ref{tab:runtime-deltas-bank}.

\begin{table}[htbp]
    \caption{Runtime and energy savings (Bank topology).}
    \label{tab:runtime-deltas-bank}
    \small
    \setlength{\tabcolsep}{2pt}
    \renewcommand{\arraystretch}{1.15}
    \begin{tabular*}{\linewidth}{@{\extracolsep{\fill}}%
                              l r r r r r @{}}
        \toprule
        Topology &
        $\Delta$CPU\% &
        $\Delta$RAM\% &
        $\Delta$I/O\% &
        \shortstack{$\Delta t$ (sec)} &
        \shortstack{$\Delta E$ (\%)} \\
        \midrule
        Bank & 48.49 & 66.25 & 0 & 85 & 61.54 \\
        \bottomrule
        \end{tabular*}
\end{table}

\subsubsection{Effectiveness - Fidelity and Success}
The effectiveness metrics confirm that the Bank twin executes \emph{all} critical Abilities present in the full run.
For this topology, ASP=1.00, and it preserves 100\% of the Techniques (TCP=1.00).  
The PES=1.00, and the order similarity \(\tau=1.00\). 
The results indicate that reduction neither deletes nor reorders causal steps.  
The final objective is reached in both runs (\(\Delta\mathrm{Obj}=0\)).

These invariants demonstrate that SCyTAG prunes only the topology entities that are not affected by the attack scenario, thereby preserving attack behavioral fidelity while reducing the attack emulation cost.

\subsection{Scalability Stress Test Case Study}
To stress-test SCyTAG's scalability, we scale up the Bank topology to a large enterprise network while maintaining the same attack scenario.
The resulting topology (Bank-XL) comprises 1,471 nodes (routers, switches, servers, workstations, and wireless devices), which are obtained by adding new departments and extending existing floors and VLANs. 
This topology preserves the overall hierarchical structure of the original Bank, with significantly more endpoints and internal segments. 
To compare the framework behaviour on both topologies, the attack scenario remains the same: the adversary starts from a compromised IP camera in a branch network, pivots through the DVR by exploiting the same CVEs, and ultimately reaches the high-value DVR asset.

We deploy the Bank-XL topology on the same test environment as the other case studies.
Building a cyber twin for the Bank-XL topology in GNS3 stresses the virtualization layer: GNS3 requires approximately 2 hours and 7 minutes to instantiate all devices, and the VM's RAM usage rises from about 0.5 GB (before activation) to an average of 12.8 GB and a peak of 15 GB out of 16 GB total RAM. 
This is an average RAM utilization of \~82\% and a peak of \~94\%. 
The system operates close to the Linux kernel's out-of-memory threshold, and the GNS3 server process is eventually terminated by the \ac{OS}.
This means that a cyber twin for the Bank-XL topology as-is \emph{cannot} be generated in our test environment.

We then run SCyTAG's Attack Graph Generator on the facts of Bank-XL with the same attack scenario IRs as in the original Bank. 
The result is a small set of topology nodes used to generate a cyber twin and emulate the attack on it.
The generated cyber twin is identical to the one generated for the Bank topology, proving the correct functionality of the framework.
Table~\ref{tab:bank-xl-reduction} compares the structural reduction obtained for Bank and Bank-XL. 
While for the original Bank topology $R_H=87.37\%$ and $R_E=88.54$\%, the Bank-XL topology achieves better results: $R_H=99.18\%$ and $R_E=99.25\%$.
The AG structure and Caldera operations remain unchanged: every Technique, Ability, and step along the attack path is preserved, with $ASP=TCP=1$. 

This case study demonstrates that in enterprise environments where a full cyber twin cannot be generated using available resources, SCyTAG can efficiently generate an attack-specific cyber twin and emulate the attack scenario.

\begin{table}[t]
    \centering
    \caption{Structural reduction achieved by SCyTAG on the Bank and Bank-XL topologies. $R_H$ and $R_E$ denote the host and edge reduction ratios, respectively.}
    \label{tab:bank-xl-reduction}
    \begin{tabular}{lrrrrrr}
        \toprule
        & \multicolumn{3}{c}{\textbf{Hosts}} & \multicolumn{3}{c}{\textbf{Connections}} \\
        \cmidrule(lr){2-4} \cmidrule(l){5-7}
        \textbf{Topology} & \textbf{Full} & \textbf{Twin} & $\mathbf{R_H}$ [\%] & \textbf{Full} & \textbf{Twin} & $\mathbf{R_E}$ [\%] \\
        \midrule
        Bank      & 95   & 12 & 87.37 & 96   & 11 & 88.54 \\
        Bank-XL   & 1471 & 12 & 99.18 & 1469 & 11 & 99.25 \\
        \bottomrule
    \end{tabular}
\end{table}

\subsection{Same Attack Scenario in Different Topologies} \label{subsec:same-scenario-different-topologies}
\begin{figure}[htbp]
  \centering
  \begin{subfigure}[b]{0.59\linewidth}
    \includegraphics[width=\linewidth]{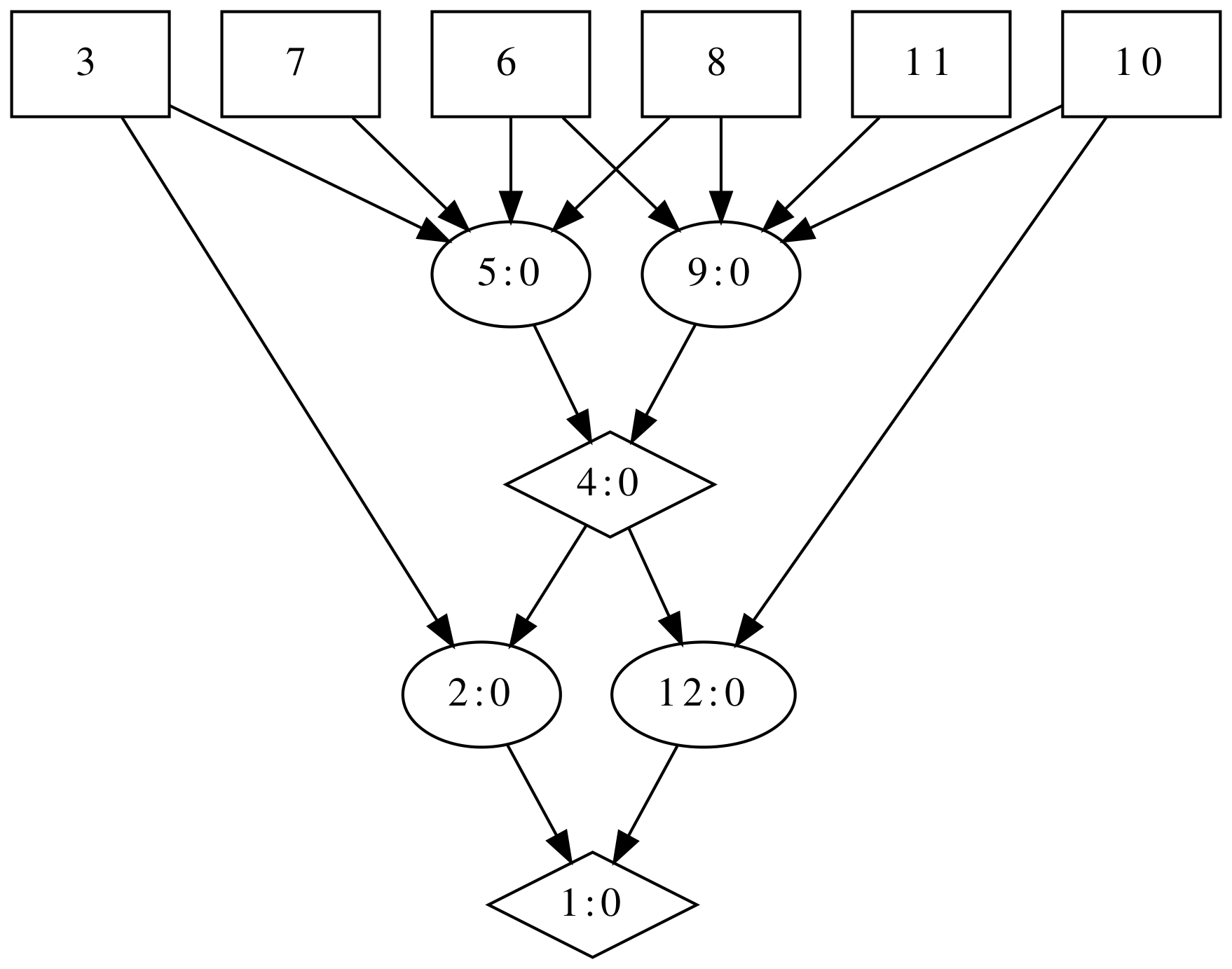}
    \caption{UK Office partial \ac{AG}.}
    \label{fig:execcode-UkOffice}
  \end{subfigure}
  \hfill
  \begin{subfigure}[b]{0.40\linewidth}
    \includegraphics[width=\linewidth]{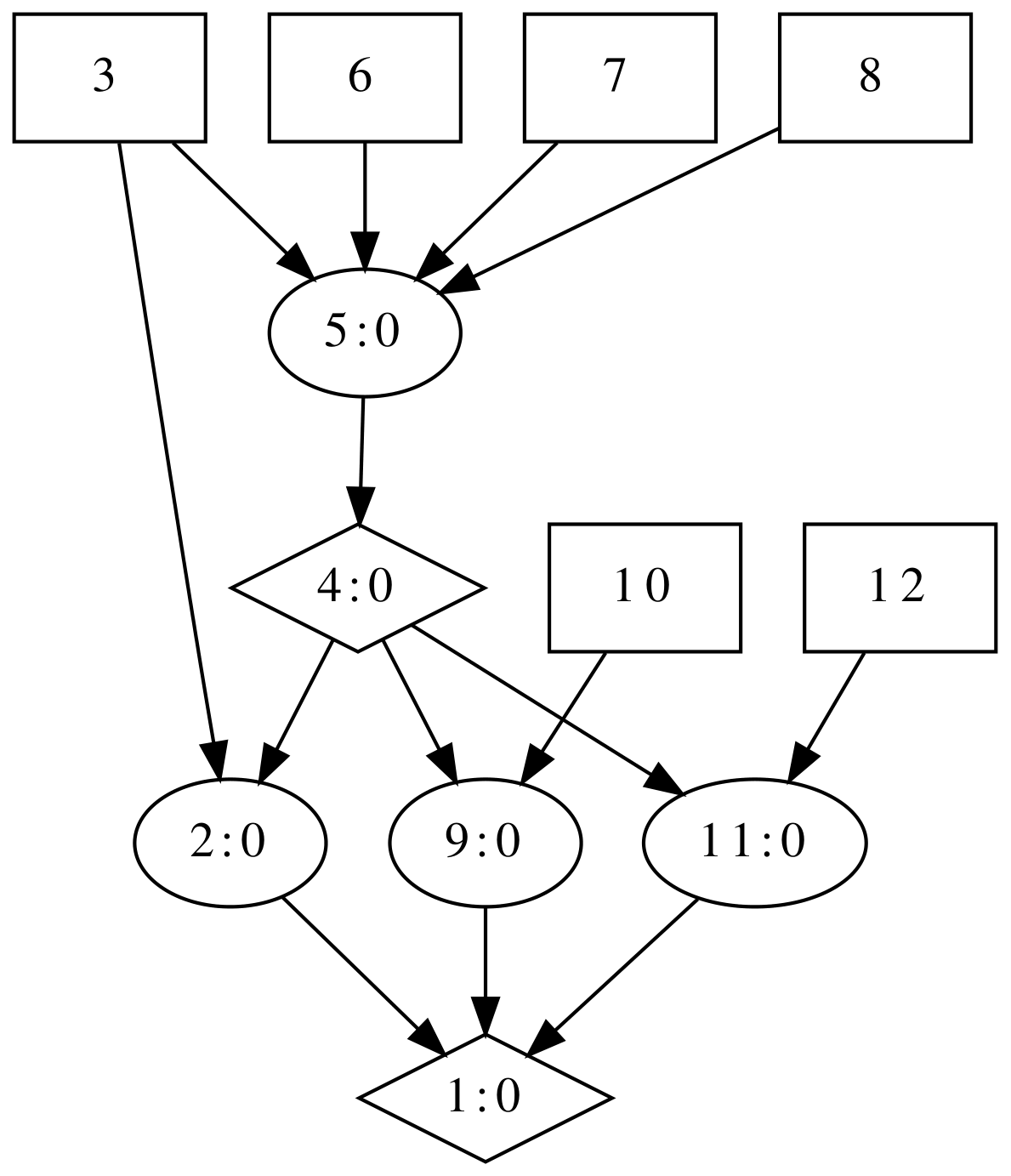}
    \caption{Bank partial \ac{AG}.}
    \label{fig:execcode-bank}
  \end{subfigure}
  \vspace{-0.5em}
  \caption{The same attack scenario performed in two topologies.}
  \label{fig:execcode}
\end{figure}

A cyber twin should be based on \emph{both} the attack logic \acp{IR} and the concrete network assets it runs on.  
To evaluate SCyTAG's ability to produce cyber twins with this capability, we run the same attack scenario in different topologies.
\Cref{fig:execcode} presents partial \acp{AG} for an attack step (\texttt{\small{execCode}}) performed in the \textbf{UK Office} and \textbf{Bank} topologies.
Both use the same \acp{IR} modeling Technique T1003 \emph{(OS Credential Dumping)}, yet the resulting \acp{AG} are different.

\begin{itemize}[leftmargin=*]
    \item When performed in the \textbf{UK Office} topology, the attack requires \emph{two} web-UI vulnerabilities on the VPN gateway (\texttt{\small{cve-2023-27524}} and a zero-day, nodes [3] and [10]).  
    The attacker must first forge a VPN certificate, obtain network reachability and local file access, and then escalate via the privileged web-UI helper.  
    The subgraph spans 12 nodes and touches three network segments before reaching \texttt{\small{execCode}}.
    \item When performed in the \textbf{Bank} topology, the attack reaches the same goal on the DVR through a different stack: 
    SMBv1 EternalBlue (\texttt{\small{cve-2017-0144}}, node [3]) plus two auxiliary \textsf{libtasn1} library flaws (nodes [10] and [12]).  
    Here, network pivoting is handled earlier in the full graph. 
    The \texttt{\small{execCode}} attack step involves only the DVR and local credential-dumping steps (9 nodes in total).
\end{itemize}

Quantitatively, the rule appears once in the UK Office topology and three times in the Bank topology, reflecting the latter’s deeper software stack.
This comparison highlights the importance of integrating both \emph{network assets} and \emph{scenario logic} when generating a cyber twin: identical \acp{IR}' conditions can be different hosts, vulnerabilities, and control paths depending on the topology in which they are executed.

\section{Related Work} \label{sec:RelatedWork}
Security-oriented \ac{DT} research has evolved along three largely independent tracks as described below.

\textbf{Risk Analytics with Attack Graphs}:
\Acp{LAG} have been embedded in \acp{DT}, primarily to rank security controls.
Kravchenko \textit{et~al.}~\cite{hadar2020cyber} connected network facts, ATT\&CK semantics, and security-control nodes in an analytical attack graph that outputs a scalar risk score.
Their prototype mirrors an enterprise of 60 hosts, yet the twin is static: no exploits are executed, and pruning is driven by control criticality, not reachability.
A dynamic \ac{LAG} that regenerates the graph whenever the host state changes, producing time-varying exposure scores, was introduced by Boudermine \textit{et al.}~\cite{boudermine2023dynamic}.
Unger \textit{et al.}~\cite{unger2023risk} embedded consequence and countermeasure nodes to create risk-assessment graphs that classify residual risk while explicitly modeling mitigations.

SCyTAG ingests external \ac{CTI}, converts its attack scenario to MulVAL rules, and builds a cyber twin based only on the hosts and flows that occur on at least one surviving path.

\textbf{High-Fidelity Twin Platforms \& CTI Generation}:
Existing networking testbeds focus on being generic experimentation tools, rather than on threat analysis.
Deng's NFV-empowered \ac{DTCP}~\cite{deng2023nfv} was shown to reproduce every router, switch, and \ac{VM} ($\sim\!10^{3}$ virtual nodes) on an eight-server OpenStack cluster to preserve topology, application, rule, and behavior fidelity.  
While ideal for fault-diagnosis studies, \ac{DTCP} lacks the ability to model threats and is hardware-intensive.  
Dietz \textit{et~al.}~\cite{dietz2022harnessing} inverted the usual pipeline: a SYN-flood is replayed against a MiniCPS conveyor-belt twin (seven ICS hosts), and logs/PCAPs are translated into a STIX~2.1 bundle.
The study showed that simulations can produce \ac{CTI}, but the simulations depend on hand-scripted attacks and a full-scope twin.

Given a \ac{CTI}, SCyTAG \emph{ingests rather than generates} threat intelligence, derives an \ac{AG}, and generates a \emph{minimal viable} cyber twin, pruning \mbox{50–85}\% of hosts/links while preserving attack behavioral fidelity.  

\textbf{Automated Adversary-Emulation Frameworks}:
Existing emulation frameworks emphasize behavioral coverage, and not environmental efficiency.  
In SpecRep~\cite{portase2024specrep}, objective graphs were compiled from white paper text, and action variants were launched on an existing lab consisting $\le\!40$~VMs. 
Lacking a reachability model, it cannot discard irrelevant hosts.  
Perry~\cite{singer2025perry} translated attacker/defender state machines into Caldera and Metasploit scripts deployed on 25-50 host OpenStack labs; an internal \ac{AG} service guides movement at runtime but never influences what is built.
Both assume that the lab already exists.  

SCyTAG differs from these methods by \emph{building and trimming} the twin \emph{before} execution.
It generates an \ac{AG}, maps each edge to Caldera Abilities, and generates a minimal cyber twin.
The limitations of existing methods:
(\emph{i}) automatic \ac{CTI} ingestion, (\emph{ii}) twin minimization, (\emph{iii}) execution of the complete TTP chain on the reduced twin, and (\emph{iv}) quantitative cost-fidelity metrics.
SCyTAG addresses all four gaps, bridging \emph{\ac{CTI} $\rightarrow$ \ac{AG} $\rightarrow$ minimal twin $\rightarrow$ real-time emulation} and enabling repeatable, cost-efficient threat assessment on large enterprise networks.

\section{Discussion} \label{sec:Discussion}
The size of the cyber twin that can be generated to emulate an attack scenario depends on the amount of allocated resources.
In our test environment (see~\cref{tab:test-env-characteristics}), we could generate a cyber twin for a network with a maximum of 575 elements.
The amount of required resources for generating a cyber twin for a network with thousands of elements will be high, making it very expensive.
SCyTAG enables the efficient building of a cyber twin with the minimal required elements and thus computing resources, in which the given attack scenario can be emulated to assess its risk to the organization.

If the emulation of the attack in the generated cyber twin is successful, defenders can infer with high probability that the attack would succeed on their network, as we ensure that all relevant IT components of the network are represented in the cyber twin.
Representing network's security elements, such as \ac{IDS} and \ac{IPS}, in the cyber twin can make it more precise, and this can be part of future work.
If the emulation of the attack fails, the defenders can infer that the attack will also fail in the network, because we take all the elements that appear in the AG generated with all the network elements, and includes all the attack paths of the given scenario.
All the network elements relevant to the attack scenario appear in the AG, and only the relevant elements appear in the AG.

ScyTAG can also work with incomplete or inaccurate inputs.
If the IT specifications or the CTI are incomplete or inaccurate, or if the results of the LLM on the inputs are not precise, they can be completed or fixed manually, or the framework will run as best effort.
If the CTI techniques are not modeled yet to MulVAL IRs, they can be modeled manually.
Automating the modeling of attack scenarios, and more specifically ATT\&CK Techniques, to IRs can be part of future work.

The primary purpose of SCyTAG is to efficiently assess the impact of a specific attack scenario on an enterprise network.
The input to the framework can be more than a single attack scenario.
Either way, it enables the analysis of cascading effects and impact analysis.
The resource allocation of the SCyTAG cyber twin for multiple attack scenarios remains efficient, as the minimal set of components is selected for each scenario, and the same components in different scenarios are represented only once in the cyber twin.
Therefore, the cyber twin includes only the relevant components, and not all the network elements.

\section{Summary and Future Work} \label{sec:Summary}
To enable the quick and efficient emulation of a threat scenario in an organization's environment, leveraging minimal resources while preserving essential network characteristics, we presented SCyTAG, a streamlined framework for continuous and optimized network threat assessment.
The framework can rapidly transform \acl{CTI} into focused, scenario-specific testing environments, effectively bridging the gap between abstract attack modeling and practical cybersecurity assessment.
By utilizing \acp{AG}, SCyTAG reduces the size of the original topology, retaining only the hosts and services required for attack scenario emulation using a cyber twin.
Evaluating SCyTAG with medium and large topologies reveals that it reduces the number of entities required for cyber twin generation by up to 99\% for large topologies, while preserving all of the required nodes for complete attack emulation.

Future research could focus on adapting the framework for other use cases, as mentioned in \cref{subsec:UseCases}.
The framework also could be applied to different environments, e.g., \ac{ICS} and the cloud.
Research could also examine the visualization of the attack emulation results for various stakeholders in the security domain and the generation of alerts based on risks identified during attack emulation.

\bibliographystyle{ieeetr}
\bibliography{References}

\begin{thebibliography}{10}

\bibitem{CybersecurityRiskAssessment}
P.~Networks, ``What is a cybersecurity risk assessment?.''

\bibitem{ThreatAssessment}
sentinelOne, ``What is threat assessment in cybersecurity?.''

\bibitem{oyama2023digital}
H.~C. Oyama, {\em Digital Twin Development and Cyberattack Resilient Optimization-Based Control Design for Process Safety, Cybersecurity, and Efficiency}.
\newblock Wayne State University, 2023.

\bibitem{sembiring2015network}
J.~Sembiring, M.~Ramadhan, Y.~S. Gondokaryono, and A.~A. Arman, ``Network security risk analysis using improved mulval bayesian attack graphs,'' {\em International Journal on Electrical Engineering and Informatics}, vol.~7, no.~4, p.~735, 2015.

\bibitem{stergiopoulos2022automatic}
G.~Stergiopoulos, P.~Dedousis, and D.~Gritzalis, ``Automatic analysis of attack graphs for risk mitigation and prioritization on large-scale and complex networks in industry 4.0,'' {\em International Journal of Information Security}, vol.~21, no.~1, pp.~37--59, 2022.

\bibitem{strom2018mitre}
B.~E. Strom, A.~Applebaum, D.~P. Miller, K.~C. Nickels, A.~G. Pennington, and C.~B. Thomas, ``Mitre att\&ck: Design and philosophy,'' {\em Technical report}, 2018.

\bibitem{tounsi2018survey}
W.~Tounsi and H.~Rais, ``A survey on technical threat intelligence in the age of sophisticated cyber attacks,'' {\em Computers \& security}, vol.~72, pp.~212--233, 2018.

\bibitem{barnum2012standardizing}
S.~Barnum, ``Standardizing cyber threat intelligence information with the structured threat information expression (stix),'' {\em Mitre Corporation}, vol.~11, pp.~1--22, 2012.

\bibitem{connolly2014trusted}
J.~Connolly, M.~Davidson, and C.~Schmidt, ``The trusted automated exchange of indicator information (taxii),'' {\em The MITRE Corporation}, pp.~1--20, 2014.

\bibitem{yara}
VirusTotal, ``Yara - the pattern matching swiss knife for malware researchers.''

\bibitem{huang2024mitretrieval}
Y.-T. Huang, R.~Vaitheeshwari, M.-C. Chen, Y.-D. Lin, R.-H. Hwang, P.-C. Lin, Y.-C. Lai, E.~H.-K. Wu, C.-H. Chen, Z.-J. Liao, {\em et~al.}, ``Mitretrieval: Retrieving mitre techniques from unstructured threat reports by fusion of deep learning and ontology,'' {\em IEEE Transactions on Network and Service Management}, 2024.

\bibitem{legoy2020automated}
V.~Legoy, M.~Caselli, C.~Seifert, and A.~Peter, ``Automated retrieval of att\&ck tactics and techniques for cyber threat reports,'' {\em arXiv preprint arXiv:2004.14322}, 2020.

\bibitem{liu2022threat}
C.~Liu, J.~Wang, and X.~Chen, ``Threat intelligence att\&ck extraction based on the attention transformer hierarchical recurrent neural network,'' {\em Applied Soft Computing}, vol.~122, p.~108826, 2022.

\bibitem{caldera}
MITRE, ``Caldera - attack emulation platform.''

\bibitem{grieves2017digital}
M.~Grieves and J.~Vickers, ``Digital twin: Mitigating unpredictable, undesirable emergent behavior in complex systems,'' {\em Transdisciplinary perspectives on complex systems: New findings and approaches}, pp.~85--113, 2017.

\bibitem{unknown}
M.~Grieves, ``Sme management forum completing the cycle: Using plm information in the sales and service functions,'' 10 2002.

\bibitem{liu2023literature}
C.~Liu, P.~Zhang, and X.~Xu, ``Literature review of digital twin technologies for civil infrastructure,'' {\em Journal of Infrastructure Intelligence and Resilience}, vol.~2, no.~3, p.~100050, 2023.

\bibitem{feng2024game}
H.~Feng, D.~Chen, H.~Lv, and Z.~Lv, ``Game theory in network security for digital twins in industry,'' {\em Digital Communications and Networks}, vol.~10, no.~4, pp.~1068--1078, 2024.

\bibitem{yu2019cybertwin}
Q.~Yu, J.~Ren, Y.~Fu, Y.~Li, and W.~Zhang, ``Cybertwin: An origin of next generation network architecture,'' {\em IEEE Wireless Communications}, vol.~26, no.~6, pp.~111--117, 2019.

\bibitem{yigit2024cyber}
Y.~Yigit, I.~Panitsas, L.~Maglaras, L.~Tassiulas, and B.~Canberk, ``Cyber-twin: digital twin-boosted autonomous attack detection for vehicular ad-hoc networks,'' in {\em ICC 2024-IEEE International Conference on Communications}, pp.~2167--2172, IEEE, 2024.

\bibitem{voas2025security}
J.~Voas, P.~Mell, P.~Laplante, and V.~Piroumian, ``Security and trust considerations for digital twin technology,'' tech. rep., National Institute of Standards and Technology, 2025.

\bibitem{ou2005logic}
X.~Ou and A.~W. Appel, {\em A logic-programming approach to network security analysis}.
\newblock Princeton University Princeton, 2005.

\bibitem{ifland2024genet}
B.~Ifland, E.~Duani, R.~Krief, M.~Ohana, A.~Zilberman, A.~Murillo, O.~Manor, O.~Lavi, H.~Kenji, A.~Shabtai, {\em et~al.}, ``Genet: A multimodal llm-based co-pilot for network topology and configuration,'' {\em arXiv preprint arXiv:2407.08249}, 2024.

\bibitem{cve}
CVE, ``Cve™ program mission.''

\bibitem{tayouri2023survey}
D.~Tayouri, N.~Baum, A.~Shabtai, and R.~Puzis, ``A survey of mulval extensions and their attack scenarios coverage,'' {\em IEEE Access}, pp.~1--1, 2023.

\bibitem{mulval2mitre}
D.~Tayouri, N.~Baum, A.~Shabtai, and R.~Puzis, ``Mulval irs mapped to mitre att\&ck tenchniques.''

\bibitem{gonda2017scalable}
T.~Gonda, R.~Puzis, and B.~Shapira, ``Scalable attack path finding for increased security,'' in {\em Cyber Security Cryptography and Machine Learning: First International Conference, CSCML 2017, Beer-Sheva, Israel, June 29-30, 2017, Proceedings 1}, pp.~234--249, Springer, 2017.

\bibitem{d3fend}
MITRE, ``D3fend™ - a knowledge graph of cybersecurity countermeasures.''

\bibitem{hadar2020cyber}
E.~Hadar, D.~Kravchenko, and A.~Basovskiy, ``Cyber digital twin simulator for automatic gathering and prioritization of security controls’ requirements,'' in {\em 2020 IEEE 28th International Requirements Engineering Conference (RE)}, pp.~250--259, IEEE, 2020.

\bibitem{boudermine2023dynamic}
A.~Boudermine, R.~Khatoun, and J.-H. Choyer, ``Dynamic logic-based attack graph for risk assessment in complex computer systems,'' {\em Computer Networks}, vol.~228, p.~109730, 2023.

\bibitem{unger2023risk}
S.~Unger, E.~Arzoglou, M.~Heinrich, D.~Scheuermann, and S.~Katzenbeisser, ``Risk assessment graphs: Utilizing attack graphs for risk assessment,'' {\em arXiv preprint arXiv:2307.14114}, 2023.

\bibitem{deng2023nfv}
L.~Deng, X.~Wei, Y.~Gao, G.~Cheng, L.~Liu, and M.~Chen, ``Nfv-empowered digital twin cyber platform: Architecture, prototype, and a use case,'' {\em Computer Communications}, vol.~210, pp.~163--173, 2023.

\bibitem{dietz2022harnessing}
M.~Dietz, D.~Schlette, and G.~Pernul, ``Harnessing digital twin security simulations for systematic cyber threat intelligence,'' in {\em 2022 IEEE 46th Annual Computers, Software, and Applications Conference (COMPSAC)}, pp.~789--797, IEEE, 2022.

\bibitem{portase2024specrep}
R.~M. Portase, A.~Colesa, and G.~Sebestyen, ``Specrep: Adversary emulation based on attack objective specification in heterogeneous infrastructures,'' {\em Sensors}, vol.~24, no.~17, p.~5601, 2024.

\bibitem{singer2025perry}
B.~Singer, Y.~Saquib, L.~Bauer, and V.~Sekar, ``Perry: A high-level framework for accelerating cyber deception experimentation,'' {\em arXiv preprint arXiv:2506.20770}, 2025.

\end{thebibliography}

\appendices
\section{Bank Topology AG}
\section{Acronyms}
\begin{acronym}
    \acro{ACL}{access control list}
    \acro{AG}{attack graph}
    \acro{API}{application programming interface}
    \acro{ASP}{ability success parity}
    \acro{CTI}{cyber threat intelligence}
    \acro{CVE}{Common Vulnerabilities and Exposures}
    \acro{DT}{digital twin}
    \acro{DTCP}{digital twin cyber platform}
    \acro{DVR}{digital video recorder}
    \acro{ICS}{industrial control system}
    \acro{IDS}{intrusion detection system}
    \acro{IPS}{intrusion prevention system}
    \acro{IR}{interaction rule}
    \acro{IT}{information technology}
    \acro{LAG}{logical attack graph}
    \acro{LLM}{large language model}
    \acro{MITM}{man in the middle}
    \acro{MTTR}{mean time to recovery}
    \acro{NLP}{natural language processing}
    \acro{OS}{operating system}
    \acro{PES}{path equivalence score}
    \acro{RAG}{retrieval augmented generation}
    \acro{SOC}{security operation center}
    \acro{TCP}{technique coverage parity}
    \acro{TTP}{tactic, technique, and procedure}
    \acro{VLAN}{virtual local area network}
    \acro{VM}{virtual machine}
\end{acronym}

\end{document}